# Exploring Damping Effect of Inner Control Loops for Grid-Forming VSCs

Liang Zhao, *Student Member*, *IEEE*, Xiongfei Wang, *Fellow*, *IEEE*, Zheming Jin, *Member*, *IEEE*

***Abstract*—This paper presents an analytical approach to explore the damping effect of inner loops on grid-forming converters. First, an impedance model is proposed to characterize the behaviors of inner loops, thereby illustrating their influence on output impedance shaping. Then, based on the impedance representation, the complex torque coefficient method is employed to assess the contribution of inner loops to system damping. The interactions among inner loops, outer loops, and the ac grid are analyzed. It reveals that inner loops shape the electrical damping torque coefficient and consequently influence both synchronous and sub-synchronous oscillation modes. The virtual admittance and current control-based inner-loop scheme is employed to illustrate the proposed analytical approach. The case study comprises the analysis of impedance profiles, the analysis of damping torque contributed by inner loops under various grid strengths, and the comparison between *dq*-frame and *αβ*-frame realizations of inner loops. Finally, simulation and experimental tests collaborate with theoretical approaches and findings.**

***Index Terms*—Voltage-source converters, grid-forming control, inner loops, virtual admittance, current control, active damping.**

## I. Introduction

Voltage-source converters (VSCs) are extensively employed with renewable energy sources, serving as an efficient interface to power grids. The grid-forming (GFM) control emerges as a favorable approach for VSC-based resources [1]. Differing from traditional grid-following control that operates VSCs to follow the frequency of grid voltage [2], GFM-VSCs behave as a voltage source behind an impedance, and they autonomously regulate the frequency and magnitude of the voltage source [3].

The control system of a GFM-VSC typically comprises outer and inner loops. The outer loop employs synchronization and voltage magnitude control to generate the electromotive force (EMF) reference vector. The synchronization control is attained by regulating their active power output or their dc-link voltage [4]-[17]. A basic approach is the active power-frequency (*P*-*f*) droop control [4], also known as the power-synchronization control (PSC) [5]. This approach can provide damping and droop characteristics but lacks the inertial response. To furnish the inertial response, the virtual synchronous generator control is developed [6], [7], which is mathematically equivalent to the *P*-*f* droop control with a low-pass filter (LPF) [8]. Besides the LPF, various forms of active power-frequency controllers are reported [9]-[11], such as the proportional-integral (PI), the P-derivative control, and the lead-lag controller, where the lead-lag controller can achieve flexible regulation of damping, inertia, and droop characteristics [11]. When dealing with regulated dc-link dynamics in GFM-VSCs, the dc-link voltage control can synthesize the frequency directly. Such dc-link voltage-synchronization control was developed for voltage angle control in wind turbines [12]. Then, similar control architectures are reported in [13]-[15]. While simple, this method may be subject to low-frequency instability issues when connected to dc-link constant-power loads [16]. To provide an additional degree of control for active damping, the dc-link voltage control and the PSC can be configured in cascade and parallel [5], [16], creating a flexible synchronization control [17]. Moreover, the magnitude of EMF reference is commonly implemented through the reactive power-voltage (*Q*-*V*) droop control [5], or the ac-bus voltage control (AVC) at the point of common coupling (PCC) [18].

Besides the outer loops, the inner loops of GFM-VSCs are of important concern. Some GFM control schemes omit the use of inner loops [5], [19], where the EMF reference vector directly serves as the modulation-voltage vector. GFM-VSCs with this open-loop voltage control have a simpler controller design, while the overload ride-through issues under large disturbances may arise [20], [21]. Virtual impedance control is a straightforward method to regulate the output impedance under both steady state and transient operations [5], [22]-[24], which is a breed of current control. The virtual inductance or the quasi-static virtual reactance can be adopted [23], while the virtual inductance emulation involves derivative calculations. An alternate inner-loop scheme is the dual-loop vector voltage and current control [25]-[30]. This method has capabilities of reference tracking, current limiting, and harmonics rejection, while it poses a risk of underdamping or even instability when interacting with synchronization control and stiff ac grid [25], [26]. Instead, employing the grid-side current disturbance feedforward can enhance the damping effect for the stability robustness [27], [28]. Moreover, an asymmetrical vector voltage control is proposed in [29] for the damping enhancement, which introduces an integral control loop from *d*-axis voltage ($v_d$) to *q*-axis current reference ($i_{qref}$) to the vector voltage controller [30]. Yet, a sophisticated parameter design is required, and the damping enhancement effect is not significant under a weak grid [30]. Moreover, the virtual admittance and current control serve as another variant of inner loops [31]. This method can emulate an inductance without derivative calculations. Yet, the voltage drop on the virtual admittance reduces the PCC voltage, and the outer-loop AVC is thus required to maintain the PCC voltage magnitude around 1p.u. [32]. A recent development involves a unified voltage control [33], which synthesizes those different voltage control methods through a universal structure with varying virtual impedance specifications. However, despite the diversity of reported inner-loop schemes, there is little discussion concerning the characterization of their dynamic behaviors.

The interactions between inner and outer loops mainly result in two types of oscillations: synchronous oscillation (SO) and sub-synchronous oscillation (SSO) [2], [34]. The SO issue originates from the dynamics of the power-angle control plant [5]. The inductive power transfer impedance introduces a pair of conjugate poles at the synchronous frequency, i.e., the fundamental frequency of the system. The basic principle of mitigating the SO issue is to increase the resistive part of the power transfer impedance between the internal-voltage phasor and the grid-voltage phasor. The inner control loops shape the output impedance of GFM-VSCs, thereby affecting the SO mode. The specific control loops for increasing the resistive effect include the virtual impedance control [5], the virtual admittance control [34], and the current control [34].

The SSO is another oscillatory mode, manifesting as oscillations below the fundamental frequency [24], [25]. For the virtual impedance control-based inner loop, a large virtual renaissance with a high-pass filter (HPF) may jeopardize the SSO, while utilizing a virtual reactance can mitigate it [24]. As for the dual-loop vector voltage and current control, the interaction between inner-loop vector voltage control and outer-loop PSC under a stiff grid may induce the SSO [25]. Increasing the voltage control bandwidth and employing an asymmetrical voltage controller can alleviate such adverse sub-synchronous control interactions [30], [35]. Moreover, when the virtual admittance and current control serve as the inner loop, the effects of inner-loop parameters on the SSO-mode damping ratio are investigated in [34] and [36] via sensitivity analysis. However, those analysis results are based on numerical methods, which cannot provide analytical or physical insight into interactions and instability issues. In [35], the effect of dual-loop vector voltage and current control is analyzed from the damping torque perspective. Yet, a quantitative assessment of damping is still missing. Therefore, there is still a lack of analytical assessment for the damping contribution of inner loops of GFM-VSCs.

In this paper, an analytical approach is developed to investigate the damping effect of inner loops. It comprises the impedance-based dynamics characterization and the complex torque coefficient-based control interaction analysis. First, an impedance model is proposed to uniformly represent the dynamics of various inner-loop schemes. The impedance-shaping effect of inner-loop configurations and parameters is analyzed. Then, based on the impedance representation of inner loops, the complex torque coefficient method is utilized to reveal the interactions among outer loops, inner loops, and the grid impedance. The electrical damping torque quantifies the contributions of inner loops to the damping effect on both SO and SSO modes. Following this, the proposed analytical method is exemplified using the virtual admittance and current control-based inner loops. Parameter tuning guidelines are subsequently developed to enhance the system stability robustness. Moreover, the *dq*- and *αβ*-frame realizations of inner loops are compared in terms of the damping effect. The theoretical method and analytical findings are finally validated by simulation and experimental results.

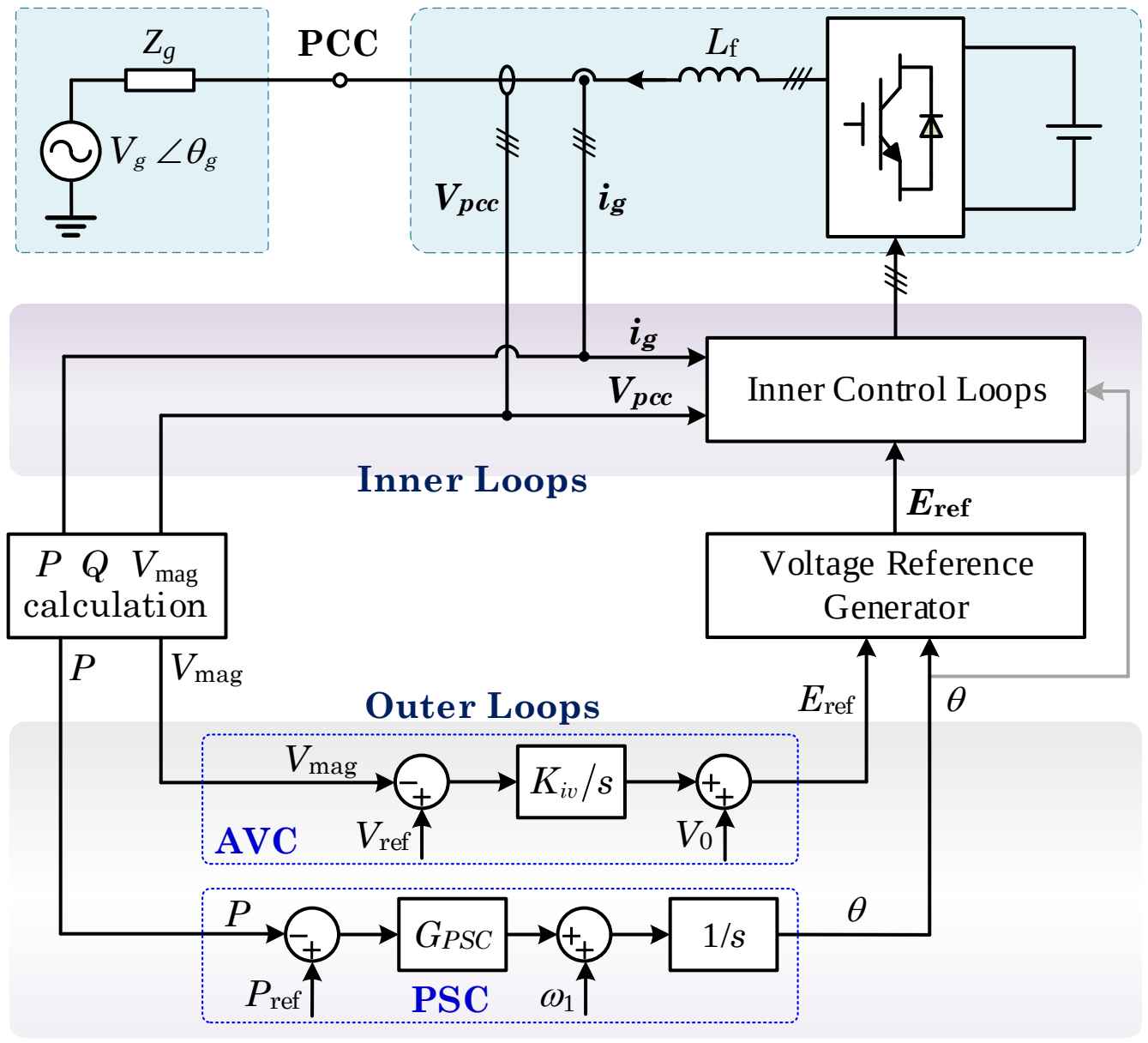

Fig. 1. System and control block diagrams of GFM-VSCs.

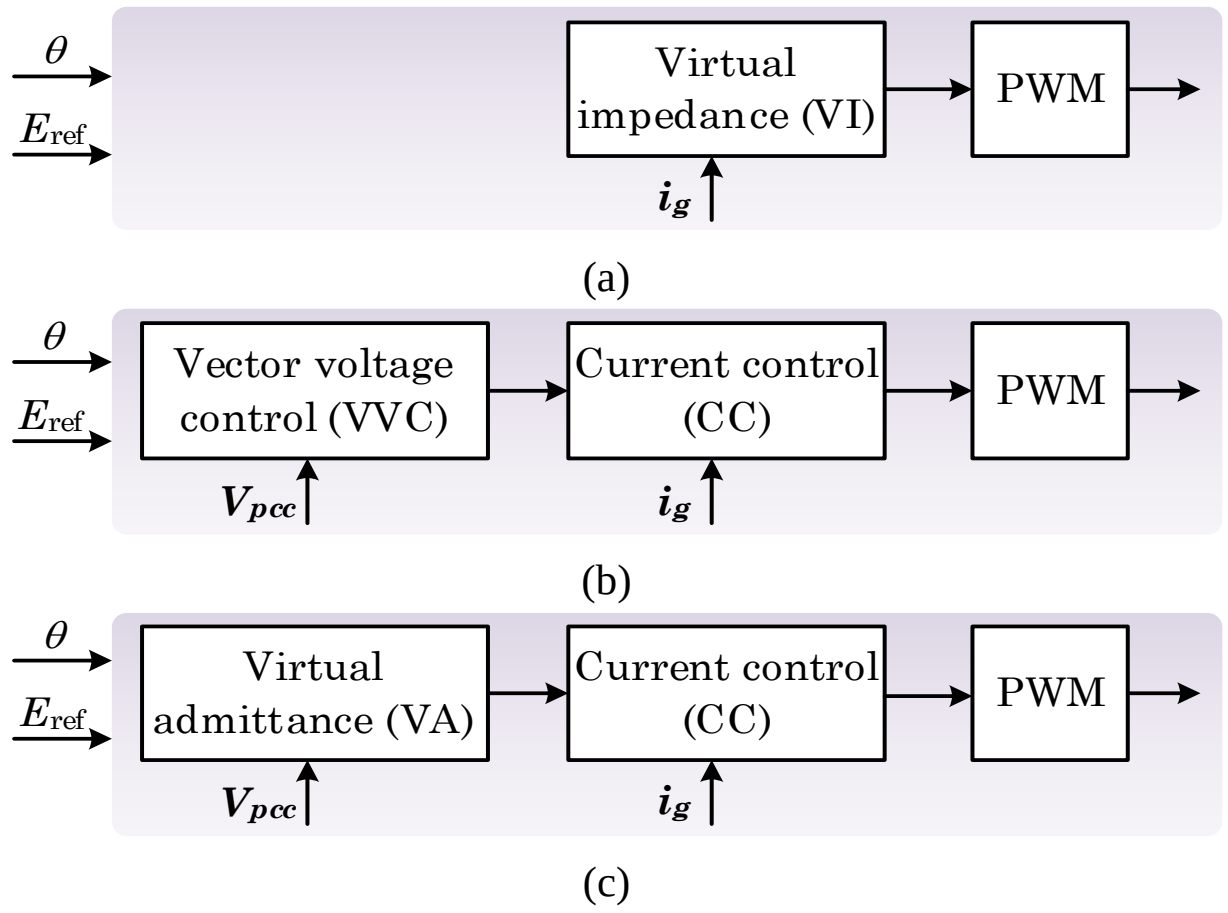

Fig. 2. Three types of inner-loop configurations. (a) Virtual impedance control. (b) Vector voltage and current control. (c) Virtual admittance and current control.

## II. System Description

Fig. 1 shows the system and control block diagram of the GFM-VSC. The single-converter infinite bus system is employed for the study, where the ac grid is represented by an ideal voltage source, $\boldsymbol{V_g} = V_g \angle \theta_g$ , in series with the grid impedance $Z_g$ [37]. Such a setting is dedicated to analyzing the stability of a single converter, while it cannot cover the interaction issues of multiple converters. The VSC is connected to the PCC through the inductor filter $L_f$. $\boldsymbol{V_{pcc}}$ and $\boldsymbol{i_g}$ denote the PCC voltage and grid current, respectively.

The control scheme of GFM-VSC comprises of outer and inner loops. The outer loops generate the EMF-vector reference, including the phase angle and magnitude. The PSC is employed for the synchronization control, where $G_{PSC}$ denotes the controller, $P$ and $P_{\mathrm{ref}}$ denote the active power and its reference, respectively. The magnitude of the EMF vector is regulated by the AVC, which is intended to regulate the PCC voltage magnitude with zero steady-state error [5]. $V_{\mathrm{mag}}$ and $V_{\mathrm{ref}}$ denote the PCC voltage magnitude and its reference, respectively. $K_{iv}$ denotes the integral gain of AVC.

Fig. 2 shows three types of inner-loop configurations that are widely used in GFM-VSCs.

1) Virtual impedance control, as shown in Fig. 2(a).
2) Dual-loop vector voltage and current control, as shown in Fig. 2(b).
3) Virtual admittance and current control, as shown in Fig. 2(c).

## III. Impedance-Based Dynamic Characterization

This Section employs an impedance model to characterize the dynamics of inner loops. The impedance shaping effect of inner loops is analyzed, which utilizes the virtual admittance and current control as a typical case.

### A. Basic Principle and Derivation

Fig. 3 illustrates the basic principle of the impedance-based methodology. Fig. 3(a) shows the GFM-VSC with studied inner loops. The outer lops generate the EMF-vector reference, $\boldsymbol{E}_{\mathbf{ref}}$, via the voltage reference generator. Afterward, inner loops process $\boldsymbol{E}_{\mathbf{ref}}$ and generate the modulation-voltage reference. In contrast, Fig. 3(b) shows the GFM-VSC with impedance-based representation of inner loops. The inner loops are omitted, while an impedance $\boldsymbol{Z}_{\boldsymbol{eq}}$ is utilized to represent the dynamics of inner loops. It can be observed from the comparison that inner loops result in the shaping of $L_f$ into $\boldsymbol{Z}_{\boldsymbol{eq}}$.

Fig. 4 shows the derivation process of $\boldsymbol{Z}_{\boldsymbol{eq}}$. The control block diagram and derivation are performed in the $\alpha\beta$-frame. **Stage 1** shows the initial control block diagram of inner loops. It is a uniform representation of different inner-loop configurations, as shown in Fig. 2, including the voltage control, current control, and voltage decoupling control. $F_{vc}$ denotes the flag of the voltage control, which can be zero or one. $G_v$ denotes the voltage controller, where different forms can denote the vector voltage control and virtual admittance control. $F_{cc}$ denotes the flag of current control, which can be unity gain or the virtual impedance. $G_i$ denotes the current controller. $F_v$ denotes the flag of voltage decoupling control, which can be zero or the LPF employed in voltage decoupling control. $G_d$ denotes the time delay.

**Stage 2:** According to the Block diagram algebra and the Superposition theorem [38], [39], the controlled voltage source can be decomposed into three series controller voltage sources, generated by $\boldsymbol{E}_{\mathbf{ref}}$, $\boldsymbol{V}_{\boldsymbol{pcc}}$, and $\boldsymbol{i}_{\boldsymbol{g}}$, respectively.

$$\begin{cases} \boldsymbol{U}_{\boldsymbol{i}} = \boldsymbol{U}_{\boldsymbol{i1}} + \boldsymbol{U}_{\boldsymbol{i2}} + \boldsymbol{U}_{\boldsymbol{i3}} \\ \boldsymbol{U}_{\boldsymbol{i1}} = \left(G_d \cdot G_i \cdot G_v\right) \cdot \boldsymbol{E}_{\mathbf{ref}} \\ \boldsymbol{U}_{\boldsymbol{i2}} = \left(G_d \cdot F_v - G_d \cdot G_i \cdot G_v \cdot F_{vc}\right) \cdot \boldsymbol{V}_{\boldsymbol{pcc}} \\ \boldsymbol{U}_{\boldsymbol{i3}} = \left(-G_d \cdot G_i \cdot F_{cc}\right) \cdot \boldsymbol{i}_{\boldsymbol{g}} \end{cases} \tag{1}$$

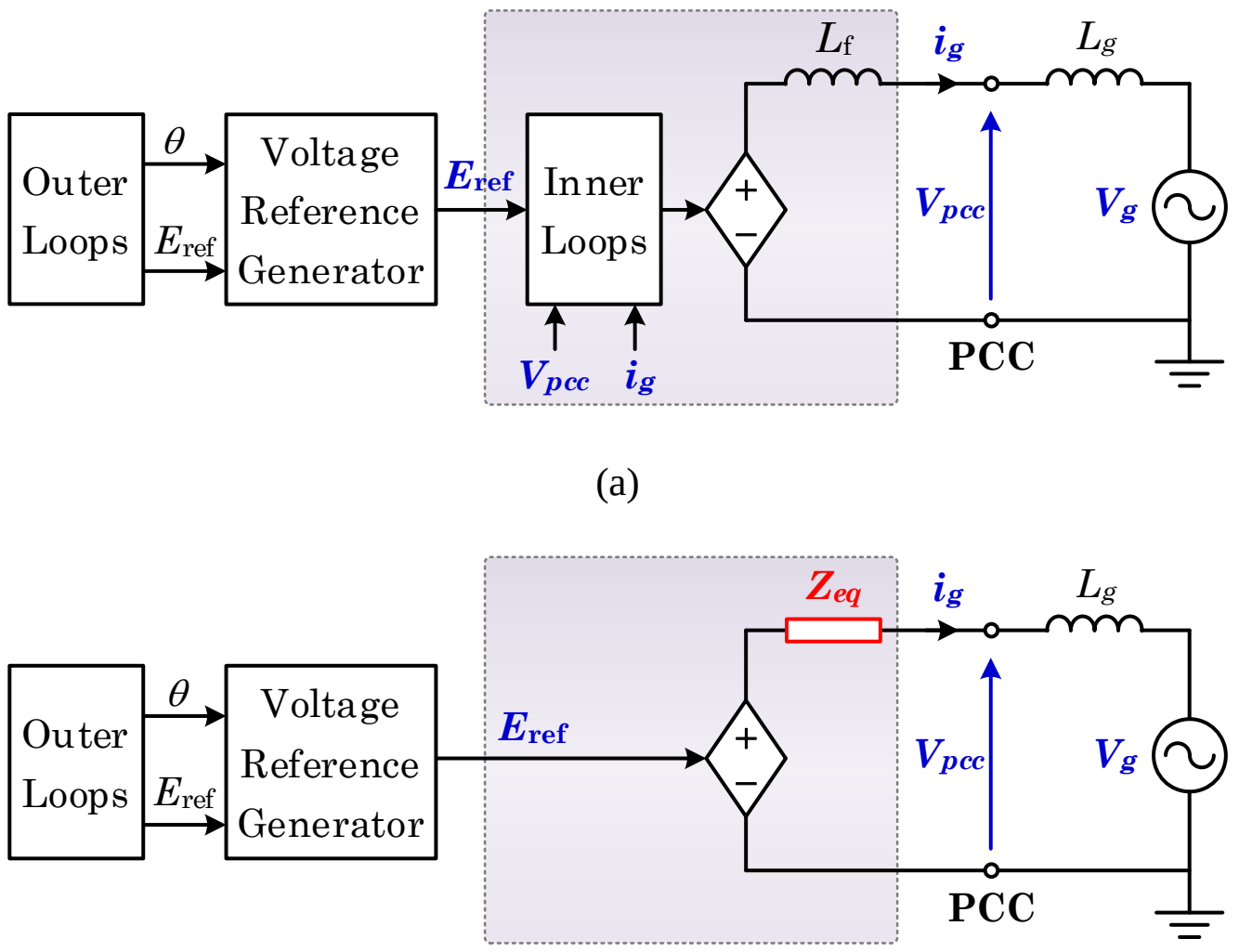

(a)

(b)

Fig. 3. Basic principle of the impedance-based method. (a) GFM-VSC with studied inner loops. (b) GFM-VSC with impedance representation of inner-loop dynamics.

**Stage 3:** According to Ohm's law [38], the current-controlled voltage source $\boldsymbol{U}_{\boldsymbol{i3}}$ can be represented with a virtual impedance, given by

$$\boldsymbol{Z}_1 = G_d \cdot G_i \cdot F_{cc} \tag{2}$$

**Stage 4:** According to Norton's theorem [38], a voltage source in series with impedances can be equivalently represented with a current source parallel with admittances.

$$\begin{cases} \boldsymbol{i}_{\boldsymbol{i1}} = \dfrac{\boldsymbol{U}_{\boldsymbol{i1}}}{\boldsymbol{Z}_1 + sL_f} = \dfrac{G_d \cdot G_i \cdot G_v}{\boldsymbol{Z}_1 + sL_f} \cdot \boldsymbol{E}_{\mathbf{ref}} \\ \boldsymbol{i}_{\boldsymbol{i2}} = \dfrac{\boldsymbol{U}_{\boldsymbol{i2}}}{\boldsymbol{Z}_1 + sL_f} = \dfrac{G_d \cdot F_v - G_d \cdot G_i \cdot G_v \cdot F_{vc}}{\boldsymbol{Z}_1 + sL_f} \cdot \boldsymbol{V}_{\boldsymbol{pcc}} \end{cases} \tag{3}$$

**Stage 5:** According to Ohm's law [38], the voltage-controlled current source $\boldsymbol{i}_{\boldsymbol{i2}}$ can be represented with a virtual admittance, given by

$$\boldsymbol{Y}_{\boldsymbol{1}} = \frac{G_d \cdot \left(G_i \cdot G_v \cdot F_{vc} - F_v\right)}{\boldsymbol{Z}_1 + sL_f}. \tag{4}$$

**Stage 6:** According to Thevenin's theorem [38], a current source in parallel with admittances can be equivalently represented with a voltage source in series with impedances. The final representation of $\boldsymbol{Z}_{\boldsymbol{eq}}$ is given by

$$\boldsymbol{Z}_{\boldsymbol{eq}} = \frac{G_d \cdot G_i \cdot F_{cc} + sL_f}{G_d \cdot \left(G_i \cdot G_v \cdot F_{vc} - F_v\right) + 1}. \tag{5}$$

The loop gain, from $\boldsymbol{E}_{\mathbf{ref}}$ to the modulation voltage, is given by

$$G_{eq} = \frac{G_d \cdot G_i \cdot G_v}{G_d \cdot \left(G_i \cdot G_v \cdot F_{vc} - F_v\right) + 1}. \tag{6}$$

## B. Case Analysis with Virtual Admittance-Current Control

The virtual admittance- and current control-based inner loop is used for analysis, which is implemented in the $\alpha\beta$-frame. In such configuration, the parameters of Fig. 4 are settled as: $F_{vc}$=1 and $F_{cc}$=1. The voltage controller denotes the virtual admittance control, given by

$$G_v = \frac{1}{sL_v + G_R \cdot R_v}. \tag{7}$$

$L_v$ and $R_v$ denote the virtual inductance and virtual resistance, respectively. In order to maintain a low R/X ratio of power transfer impedance, $R_v$ is employed with a notch filter $G_R$ at the fundamental frequency [5]. Such a notch filter becomes an HPF when the virtual admittance control is implemented in the $dq$-frame [24]. The current control employs an $\alpha\beta$-frame quasi-PR controller [40], given by

$$G_i = k_{pi} + \frac{k_{ri}\, s}{s^2 + 2\omega_r s + \omega_1^2}. \tag{8}$$

$k_{pi}$ denotes the P gain. $k_{ri}$ denotes the R gain and is equal to 0.4p.u. $\omega_{ri}$ denotes the resonant cut-off frequency, which is utilized to mitigate the sensitivity of the R controller against the variation of fundamental frequency and is equal to $2\pi\ rad/s$. Note that the controller gain at fundamental frequency is reduced by $\omega_{ri}$, while it is still large enough for reference tracking. Moreover, the timed delay can be neglected when analyzing low-frequency dynamics, i.e., $G_d$=1. Under his assumption, the loop gain $G_{eq}$ is equal to 1.

In the frequency domain, the impedance can be represented as the real and imaginary parts, denoting the resistance and the reactance, respectively. Figs. 5-7 presents the impedance profiles shaped by inner-loop virtual admittance and current control, where different parameters are tested. It is known that the resistance component can dampen the dc components (or ultra-low frequency ac components) in ac voltage and current [38]. Such issue typically manifests as fundamental-frequency oscillations in the active and reactive power, namely the SO [5]. Thus, the low-frequency resistance part of $\boldsymbol{Z_{eq}}$ is utilized to reflect the damping on the SO mode.

Fig. 5 shows the impedance profiles with different P gain of current control. $R_v$=0.1p.u. and $L_v$=0.3p.u. are adopted, and the voltage decoupling loop is disabled ($F_v$=0). It is shown that the low-frequency $R_{eq}$ increases as $k_{pi}$ increases from 0.5p.u. to 3p.u., which, however, remains below the value of $R_v$. Further, $\boldsymbol{Z_{eq}}$ is equal to the value of virtual admittance if $k_{pi}$ is infinite, given by

$$\boldsymbol{Z_{eq}}\Big|_{k_{pi}\to+\infty} = sL_v + R_v \cdot G_R, \tag{9}$$

The results indicate that the current control interacts with virtual admittance control, resulting in the negative resistance effect in the $\boldsymbol{Z_{eq}}$. Increasing the proportional gain of current control can partially alleviate such a negative resistance effect.

Fig. 6 shows the impedance profiles with and without the voltage decoupling control. The voltage decoupling control is denoted by $F_v$=0 (disabled) and $F_v$=1 (enabled). The results reveal that enabling voltage decoupling control leads to an increase in the low-frequency $R_{eq}$, approaching the value of $R_v$. Consequently, voltage decoupling control effectively mitigates the low-frequency negative resistance effect resulting from the interaction between virtual admittance and current control.

Fig. 7(a) shows the impedance profiles with different $R_v$. An increase in $R_v$ from 0.08 p.u. to 0.12 p.u. is associated with an augmentation of the low-frequency $R_{eq}$, indicating a positive damping effect. The low-frequency $R_{eq}$ is increased as $R_v$ increases from 0.08p.u. to 0.12p.u., indicating the positive damping effect. Fig. 7(b) shows the impedance profiles with different $L_v$. The $L_v$ has minimal impact on the low-frequency resistance and little effect on the SO-mode dynamics.

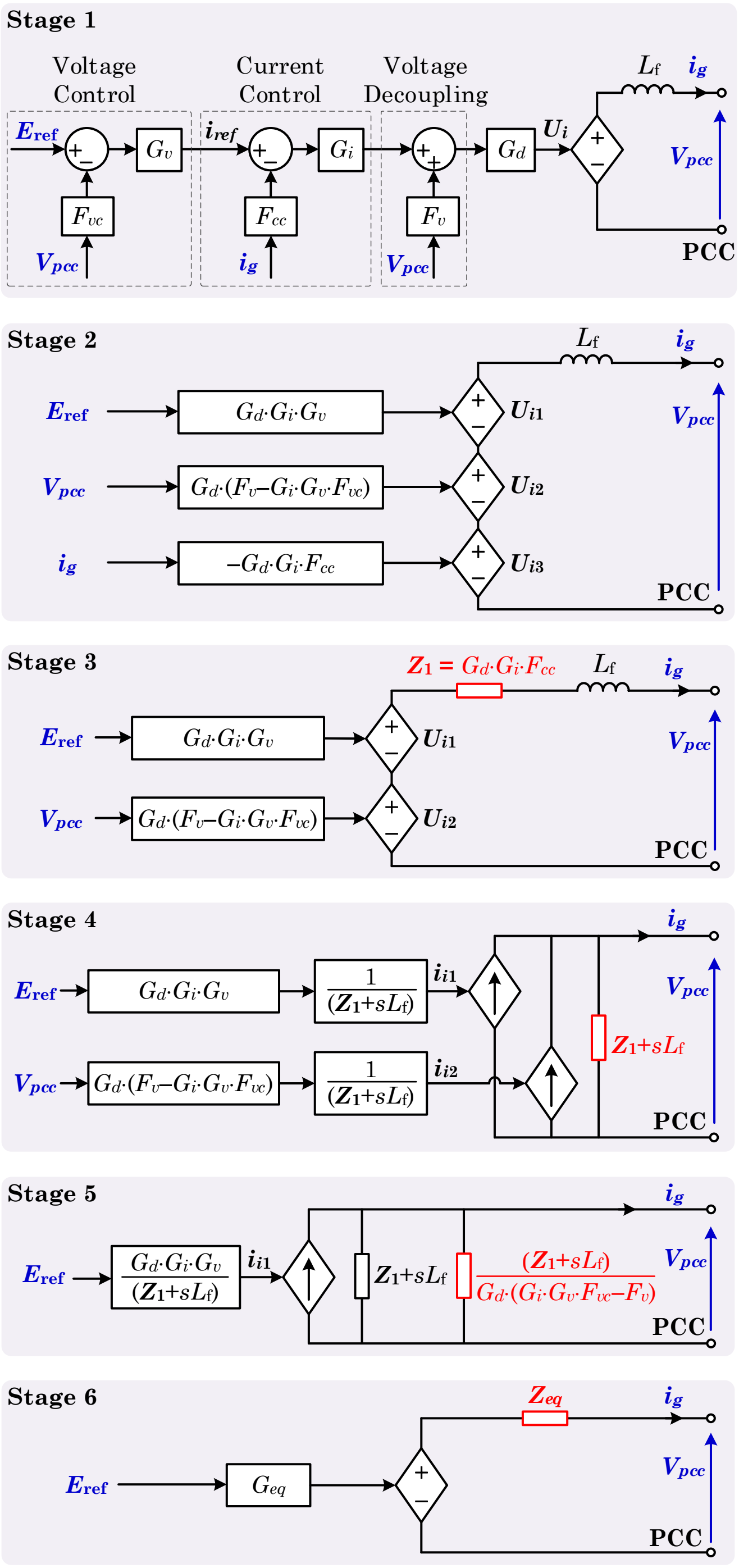

Fig. 4. Derivation of the impedance model.

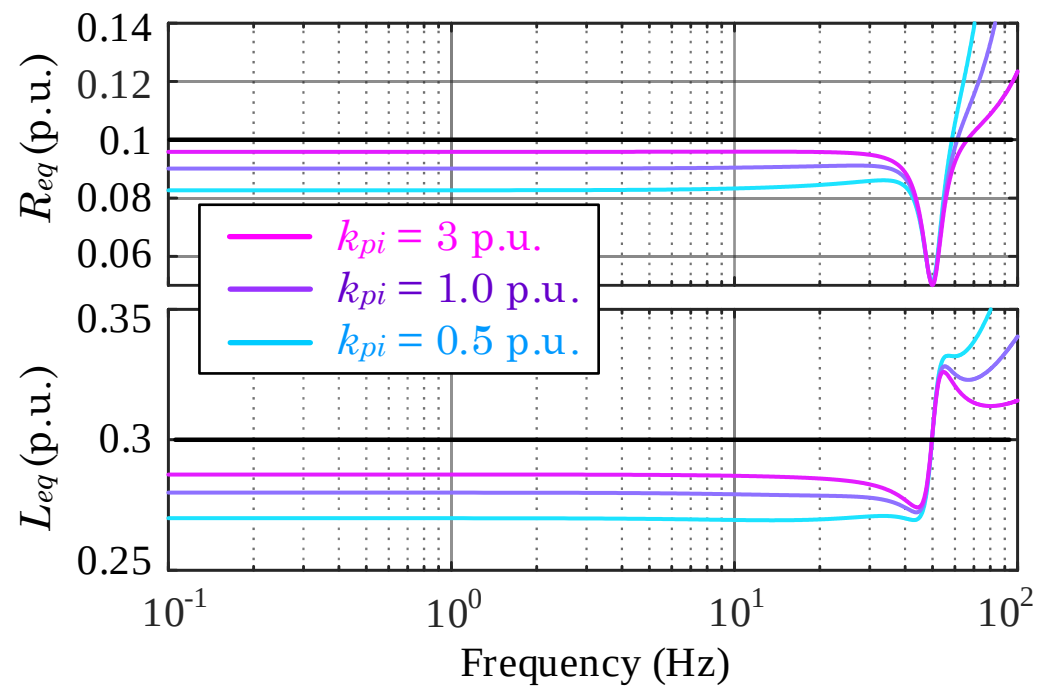

Fig. 5. Impedance profiles with different P gains of current control. $L_v$=0.3p.u., $R_v$=0.1p.u., $F_v$=0.

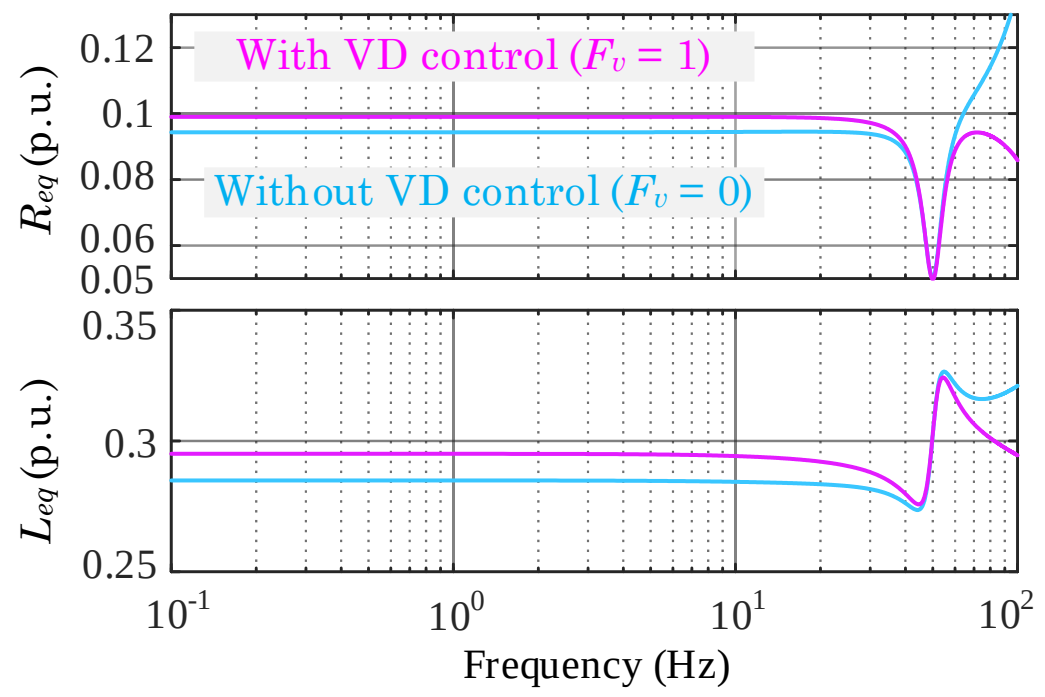

Fig. 6. Impedance profiles with and without the voltage decoupling control. $L_v$=0.3p.u., $R_v$=0.1p.u., $k_{pi}$=2p.u.

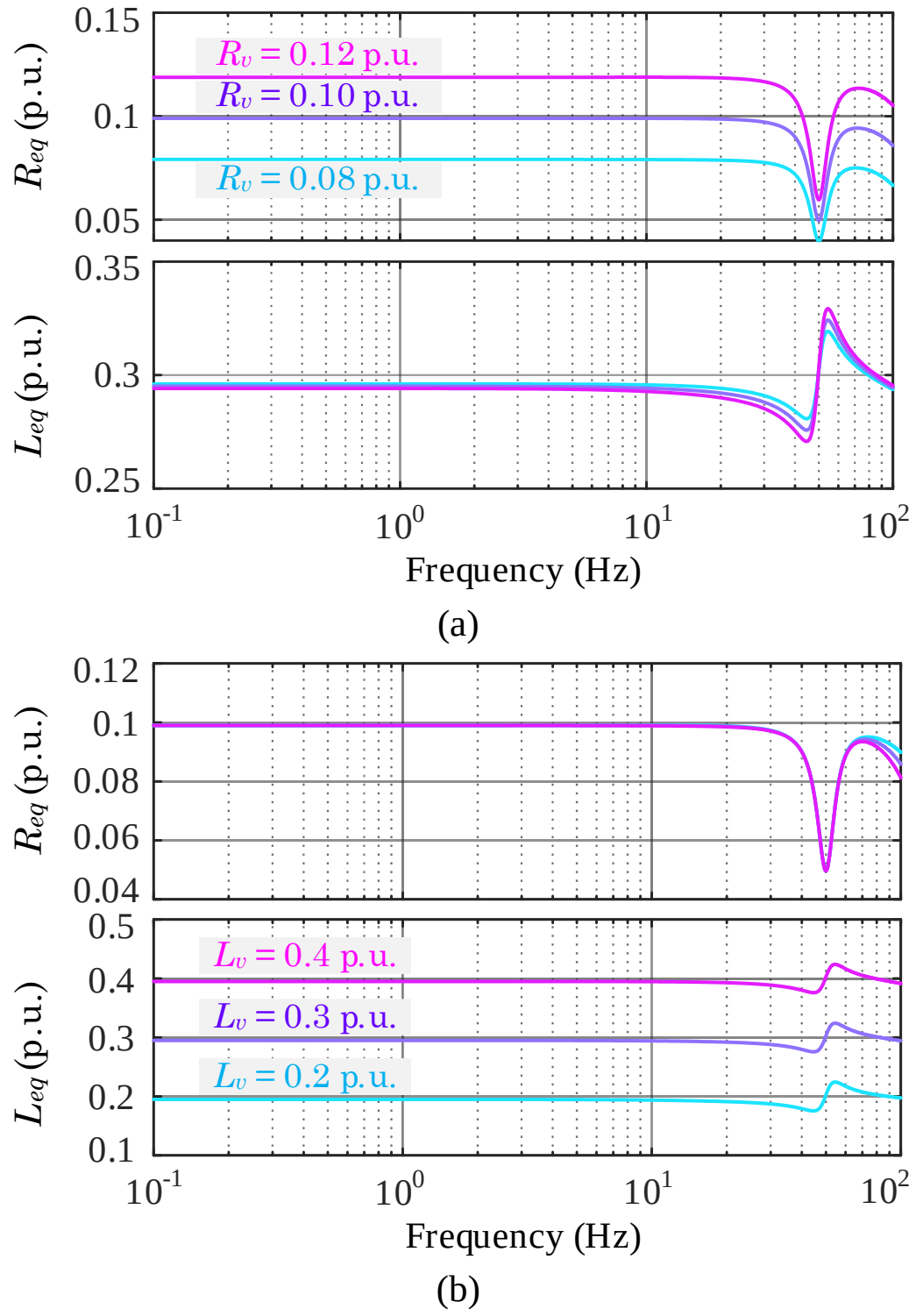

Fig. 7. Impedance profiles with different parameters of virtual admittance. $k_{pi}$=2p.u., $F_v$=1. (a) with different $R_v$. (b) with different $L_v$.

## IV. Complex Torque Coefficient-Based Interaction Analysis

This Section uses the complex torque coefficient method to reveal the interaction among inner loops, outer loops, and the ac grid. The complex torque profiles are analyzed to quantify the damping effect of inner loops on SO and SO modes.

### A. Basic Principle and Stability Criteria

The complex torque coefficient method was initially developed to investigate torsional interactions in synchronous-machine infinite-bus systems [41]. The mechanical and electrical torques characterize the dynamics of the electromechanical (rotor) and the electrical (stator) subsystems, respectively. Fig. 8 illustrates the mechanical and electrical torques in the frequency domain. The real component, aligned with the phase angle $\theta$, represents the synchronous torque. The imaginary component, aligned with frequency $\omega$, denotes the damping torque. Mathematical expressions of the complex torque are given by

$$\begin{cases} \boldsymbol{T_m}(j\omega) = \left[K_m(\omega) + j\omega D_m(\omega)\right]\cdot \Delta\theta \\ \boldsymbol{T_e}(j\omega) = \left[K_e(\omega) + j\omega D_e(\omega)\right]\cdot \Delta\theta \end{cases}. \tag{10}$$

$\boldsymbol{T_m}$ and $\boldsymbol{T_e}$ denote the mechanical and electrical torque with complex form. $K_m$ and $K_e$ represent the mechanical and electrical synchronizing torque coefficient. $D_m$ and $D_e$ denote the mechanical and electrical damping torque coefficient. Given the power/energy balance between mechanical and electrical subsystems, the equation to capture the interaction is given by

$$K_m(\omega) + K_e(\omega) + j\omega\cdot\left[D_m(\omega) + D_e(\omega)\right] = 0\,. \tag{11}$$

The real part is the sum of synchronizing torques, i.e., $K_m(\omega) + K_e(\omega) = 0$, which determines the oscillation frequencies [42]. The imaginary part is the net damping; and its value at the oscillatory frequencies determines the stability [43], i.e.

$$\begin{cases} D_m + D_e > 0 \;\Rightarrow\; \text{Stable} \\ D_m + D_e < 0 \;\Rightarrow\; \text{Unstable} \end{cases} \tag{12}$$

The advantage of this stability analysis method lies in that the contribution of mechanical and electrical subsystems to the system damping can be quantified clearly.

### B. Application of Complex Torque in GFM-VSCs

Fig. 9 shows the *dq*-frame circuit diagram with impedance representation of inner loops, which is equivalent to the *αβ*-frame circuit diagram. The *dq*-frame impedance can be derived according to the frequency translation [44], given by

$$Z_{dq}(s) = Z_{eq_\alpha\beta}(s+j\omega_1) \tag{13}$$

$$Z_{dq}(s) = Z_d(s) + jZ_q(s) \Leftrightarrow \begin{bmatrix} Z_d(s) & -Z_q(s) \\ Z_q(s) & Z_d(s) \end{bmatrix} \tag{14}$$

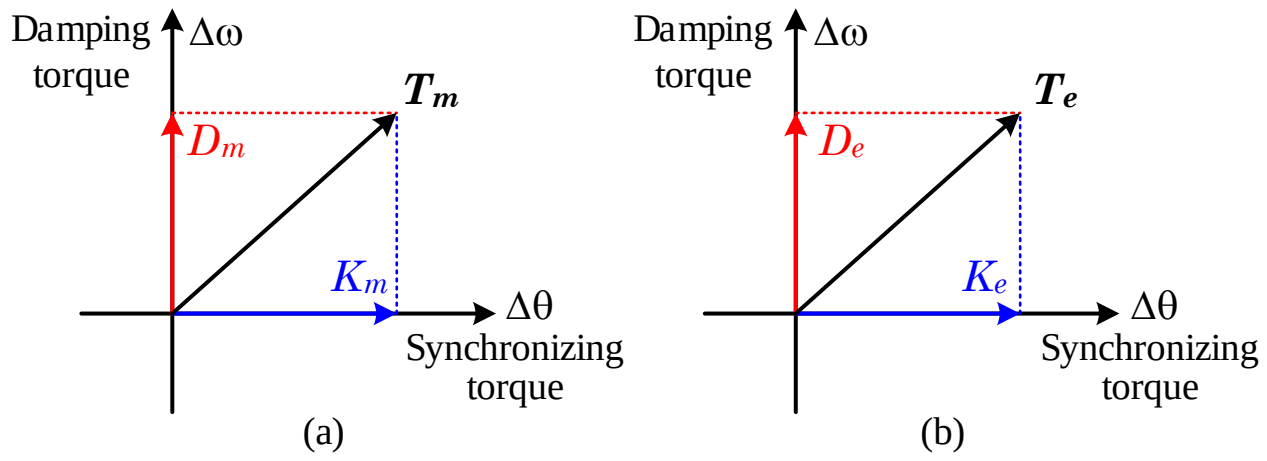

Fig. 8. Decomposition of complex torque. (a) Mechanical torque. (b) Electrical torque.

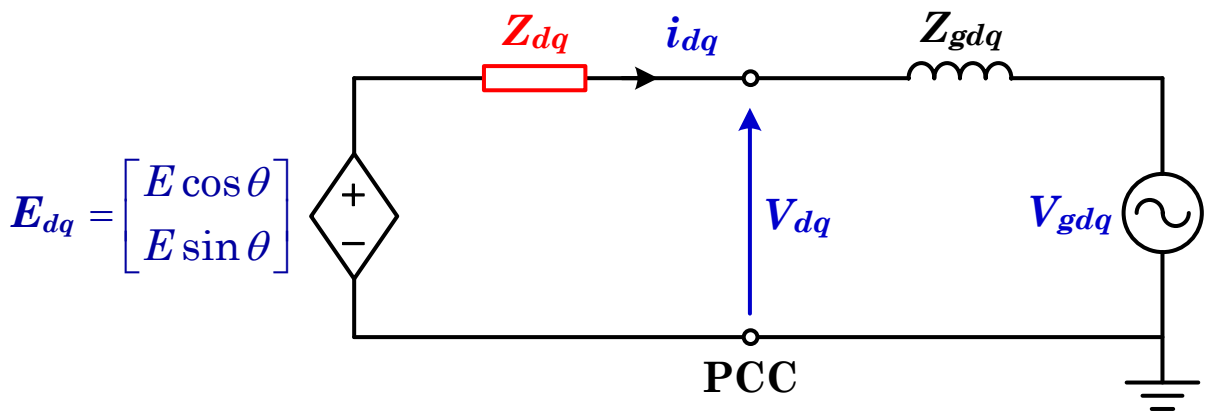

Fig. 9. Circuit diagram with impedance representation of inner loops in the *dq*-frame.

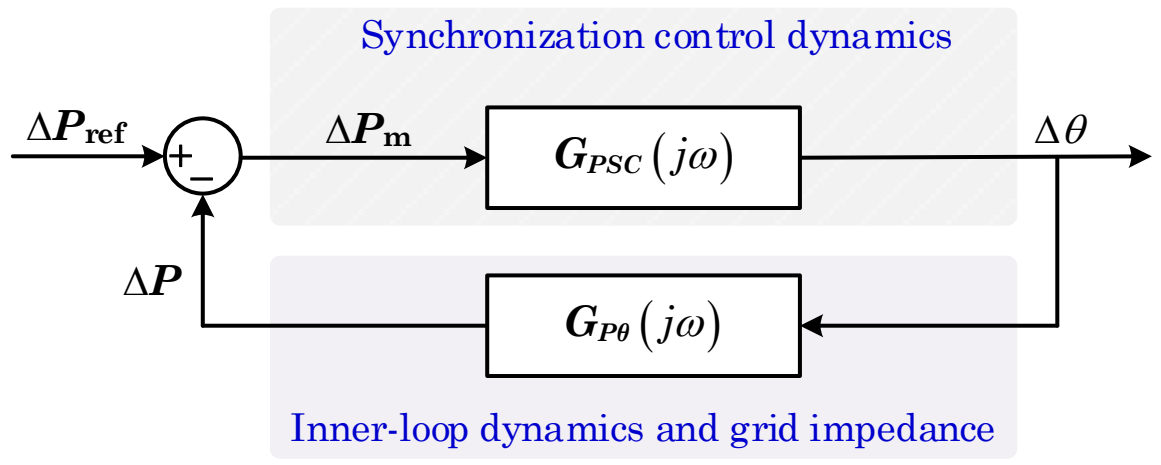

Fig. 10. Closed-loop control block diagram of the GFM-VSC.

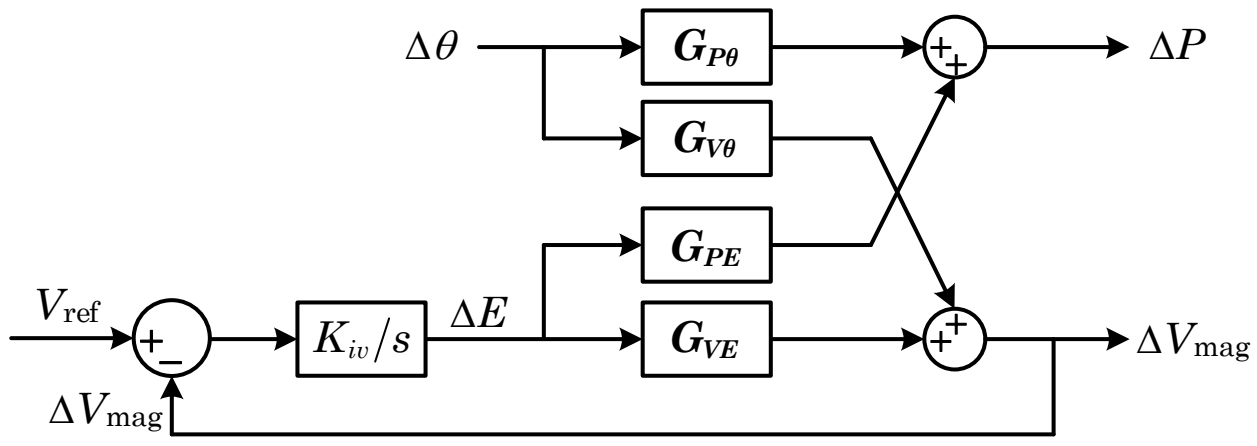

Fig. 11. Active power and voltage magnitude control plants.

TABLE I. CIRCUIT AND CONTROL PARAMETERS FOR CASE STUDIES.

| Description | Values |
|---|---|
| Grid inductance $L_g$ | Stiff grid: $L_g$=0.05 p.u. (SCR=20)<br>Weak grid: $L_g$=0.8 p.u. (SCR=1.25) |
| Outer-loop PSC:<br>$G_{PSC} = K_{psc}\cdot\omega_p/(s+\omega_p)$ | Case 1: $K_{psc}$=0.1p.u., $\omega_p$=3Hz<br>Case 2: $K_{psc}$=0.1p.u., $\omega_p$=100Hz |
| Virtual resistance $R_v$ | 0.05p.u. to 0.1p.u.<br>(With a 5Hz HPF in *dq*-frame) |
| Virtual inductance $L_v$ | 0.1p.u. to 0.3p.u. |
| Current control proportional gain $k_{pi}$ | 0.5p.u. to 2p.u. |

The active power flowing through the PCC can be calculated according to the instantaneous power theory, given by

$$P = \frac{EV_g\sin\theta(Z_q+Z_{gq}) + (E^2 - EV_g\cos\theta)(Z_d+Z_{gd})}{(Z_d+Z_{gd})^2 + (Z_q+Z_{gq})^2} \tag{15}$$

The active power versus phase angle, under small-signal perturbations, is expressed as

$$\Delta P = E_0 V_{g0}\underbrace{\frac{(Z_q+Z_{gq})\cos\theta_0 + (Z_{gd}-Z_d)\sin\theta_0}{(Z_d+Z_{gd})^2 + (Z_q+Z_{gq})^2}}_{G_{P\theta}}\cdot\Delta\theta \tag{16}$$

$E_0$, $V_{g0}$, $\theta_0$ denote the operating point of EMF magnitude, grid voltage magnitude, and the power angle. It is clear from the equation that the power-angle dynamics are affected by the inner loops, the grid impedance, and the operating point.

Fig. 10 shows the closed-loop control block diagram of the GFM-VSC in the frequency domain. $G_{PSC}$ represents the synchronization dynamics. $G_{P\theta}$ includes the inner-loop dynamics and the grid impedance, as illustrated in (16). It is clear that $1/G_{PSC}$ and $G_{P\theta}$ correspond to mechanical and electrical torque coefficients in (10), respectively. The power-angle relationship is rewritten as

$$\begin{cases} \Delta P_m = \dfrac{1}{G_{PSC}(j\omega)}\cdot\Delta\theta = [K_m(\omega) + j\omega D_m(\omega)]\cdot\Delta\theta \\ \Delta P_e = G_{P\theta}(j\omega)\cdot\Delta\theta = [K_e(\omega) + j\omega D_e(\omega)]\cdot\Delta\theta \end{cases} \tag{17}$$

The control interaction can thus be analyzed through the frequency response of $1/G_{PSC}$ and $G_{P\theta}$.

Moreover, when integrating the dynamics of AVC into the complex torque coefficient, the control plant of active power and voltage magnitude is shown in Fig. 11. The transfer functions $G_{P\theta}$, $G_{PE}$, $G_{V\theta}$, and $G_{VE}$ can be calculated through the circuit diagram Fig. 9. The transfer function from phase angle to active power can be reformulated by selecting the phase angle as the input signal and active power as the output signal, given by

$$\Delta P = \left(G_{P\theta} - G_{PE}\cdot\frac{K_{iv}/s}{1+K_{iv}/s\cdot G_{VE}}\cdot G_{V\theta}\right)\cdot\Delta\theta \tag{18}$$

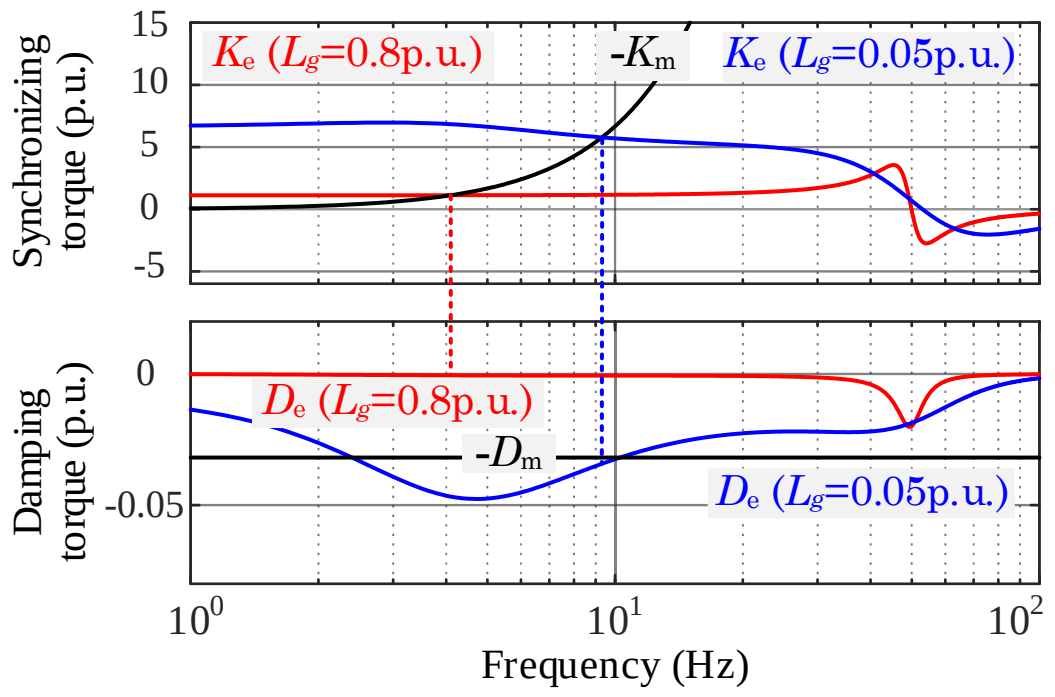

Fig. 12. Complex torque profiles with different grid impedances.

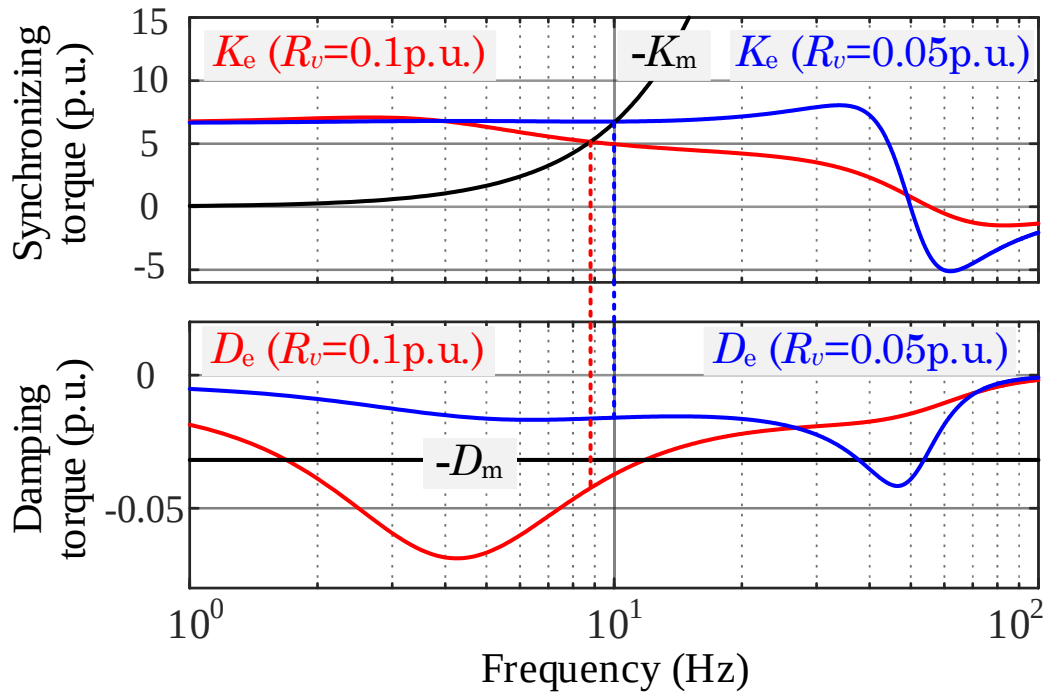

(a)

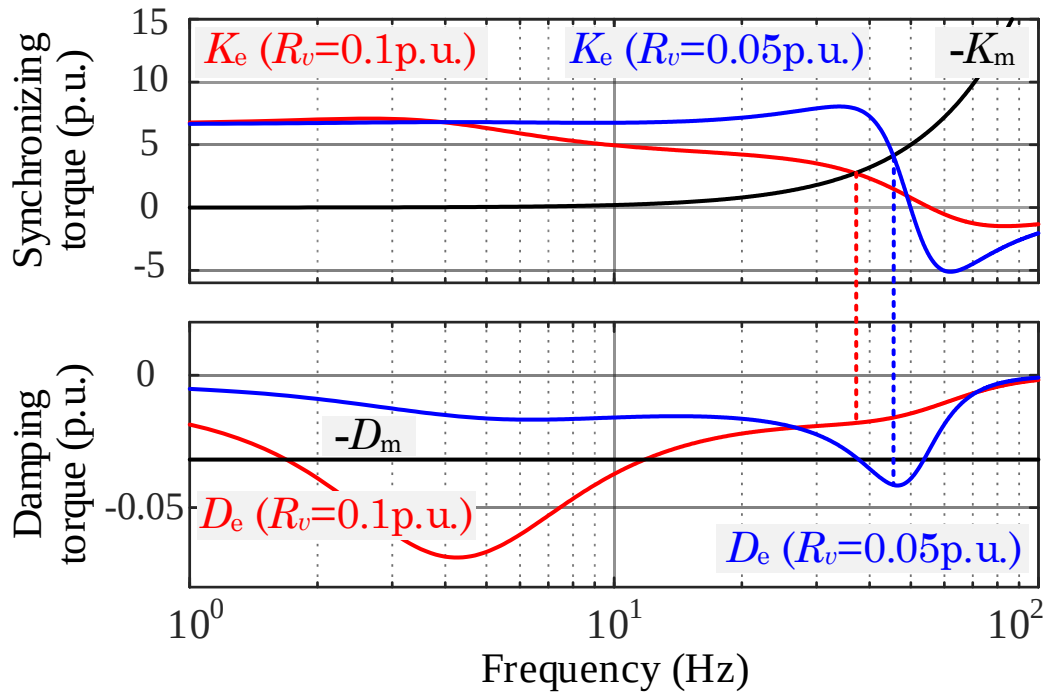

(b)

Fig. 13. Complex torque profiles with different $R_v$. (a) Outer-loop PSC with 3Hz LPF. (b) Outer-loop PSC with 100Hz LPF.

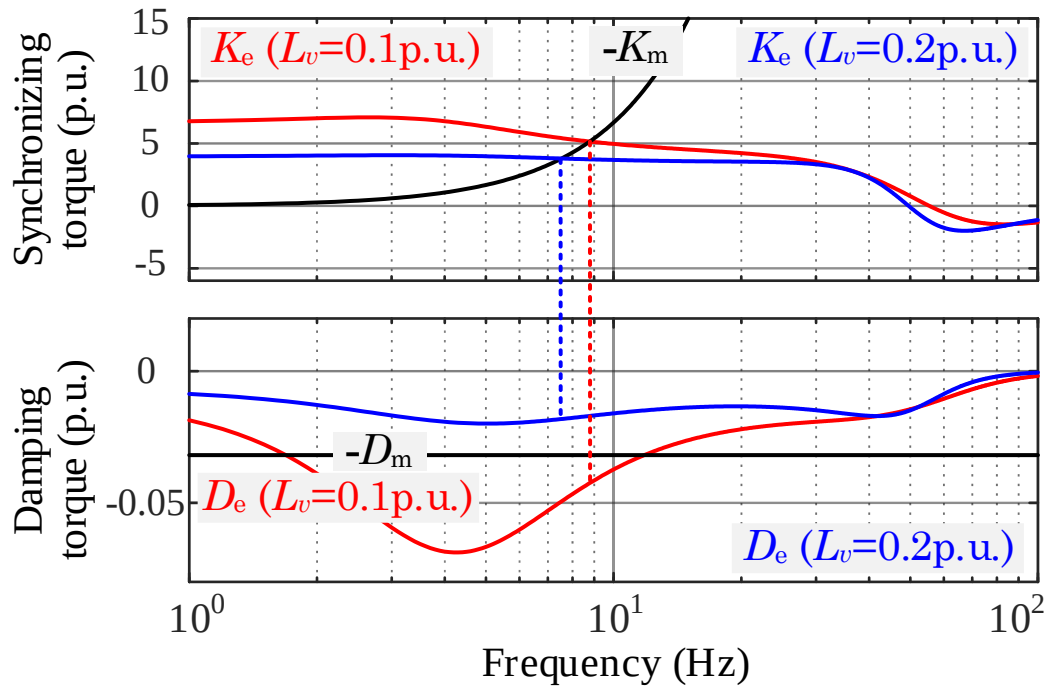

Fig. 14. Complex torque profiles with different $L_v$.

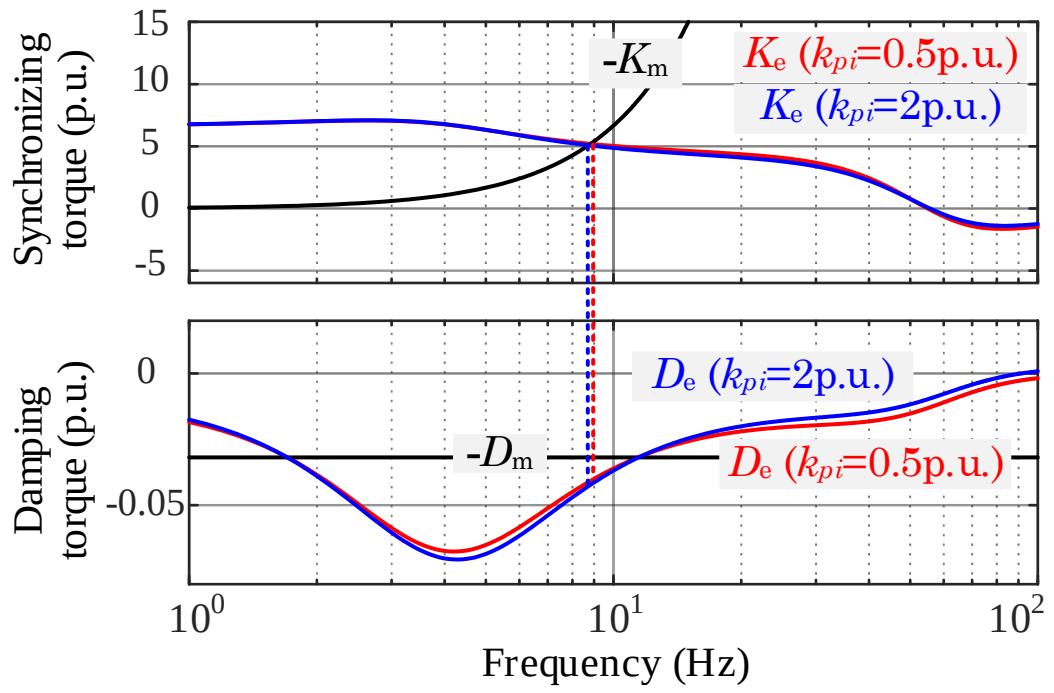

(a)

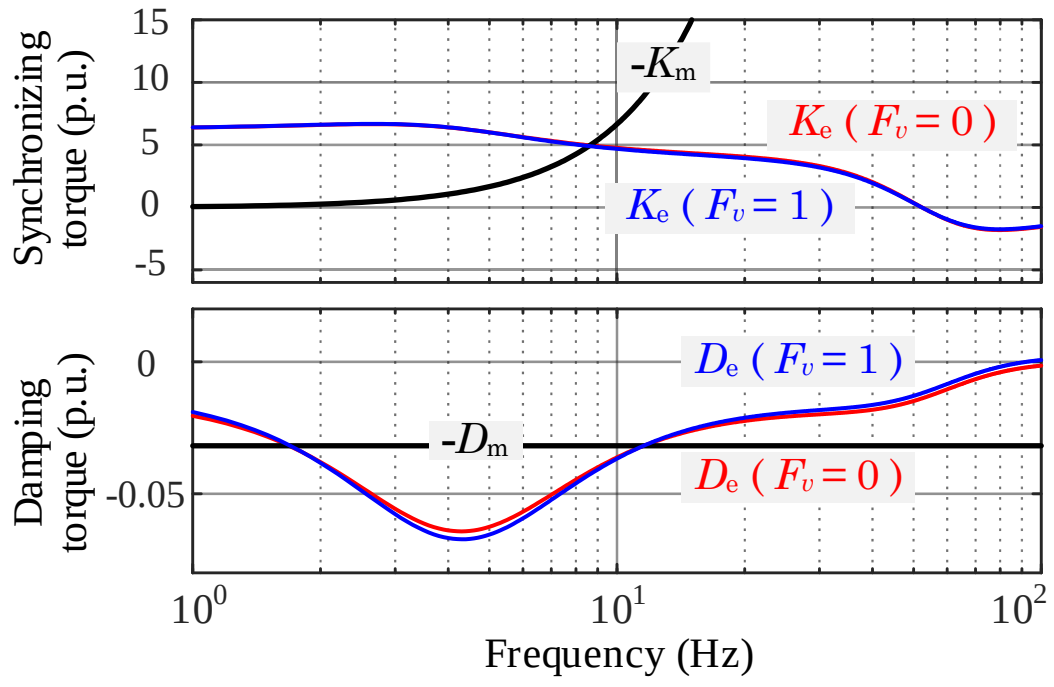

(b)

Fig. 15. Complex torque profiles with different parameters of current control and voltage decoupling control. (a) with different P gains of current control. (b) with and without the voltage decoupling control.

## C. Case Analysis with Virtual Admittance-Current Control

Table I shows the circuit and control parameters for the case study. It is noted that the 100Hz LPF in the PSC loop is employed to mitigate noises in power calculation, while 3Hz LPF is the case of inertia emulation. Additionally, the $\alpha\beta$-frame notch filter employed with $R_v$ is replaced with an HPF in the $dq$-frame [5], [24].

Fig. 12 shows complex torque profiles with different grid impedances, where $L_v$=0.1p.u., $R_v$=0.1p.u., $k_{pi}$=2p.u., and PSC employs a 3Hz LPF. The damping torque $D_e$ with $L_g$=0.8p.u. is larger than that with $L_g$=0.05p.u. It indicates that the net damping torque $D_e + D_m$ is prone to be negative under stiff grids. The results provide an analytical interpretation that the GFM-VSC exhibits a higher instability risk under stiff grids. Therefore, the following case studies are conducted under the grid condition $L_g$=0.05p.u.

Fig. 13 shows complex torque profiles with different virtual resistance $R_v$, with $L_v$=0.1p.u. and $k_{pi}$=2p.u. In Fig. 13(a), the synchronizing torques $K_e$ and $K_m$ intersect in the frequency range of the SSO mode. When $R_v$=0.05p.u., the damping torque relationship at the interaction frequency is $D_e > -D_m$, thereby leading to a positive net damping and a stable operation. In contrast, when $R_v$ changes to 0.1p.u., the damping torque relationship becomes $D_e + D_m < 0$, indicating the SSO issue. The comparative results show that increasing $R_v$ jeopardizes the damping torque of the SSO mode. In Fig. 13(b), the synchronizing torques $K_e$ and $K_m$ intersect at the fundamental frequency, referring to the SO mode. As $R_v$ increases from 0.05p.u. to 0.1p.u., the net damping torque at synchronous frequency changes from negative to positive, thereby enhancing the damping effect on the SO mode. The results in Fig. 13 demonstrate the trade-off of $R_v$ between SSO-mode and SO-mode damping torque.

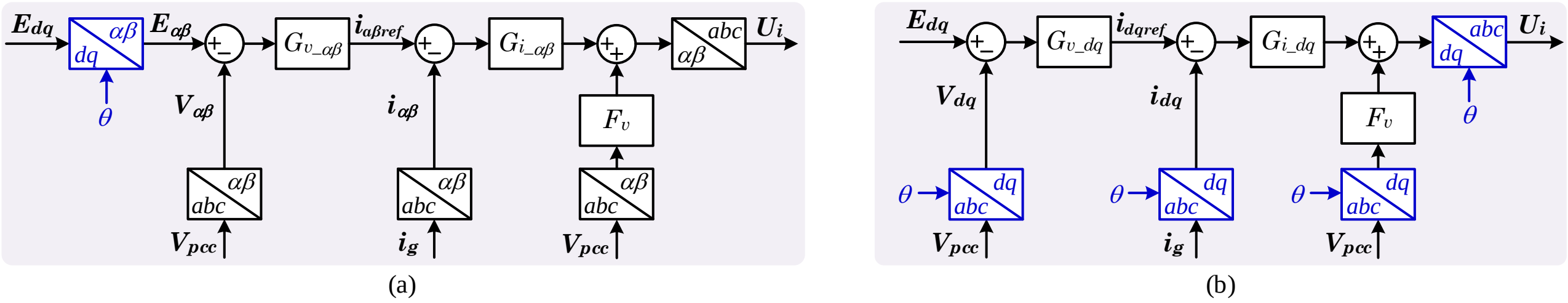

Fig. 16. Different reference frame realizations of inner loops. (a) $\alpha\beta$-frame realization. (b) $dq$-frame realization.

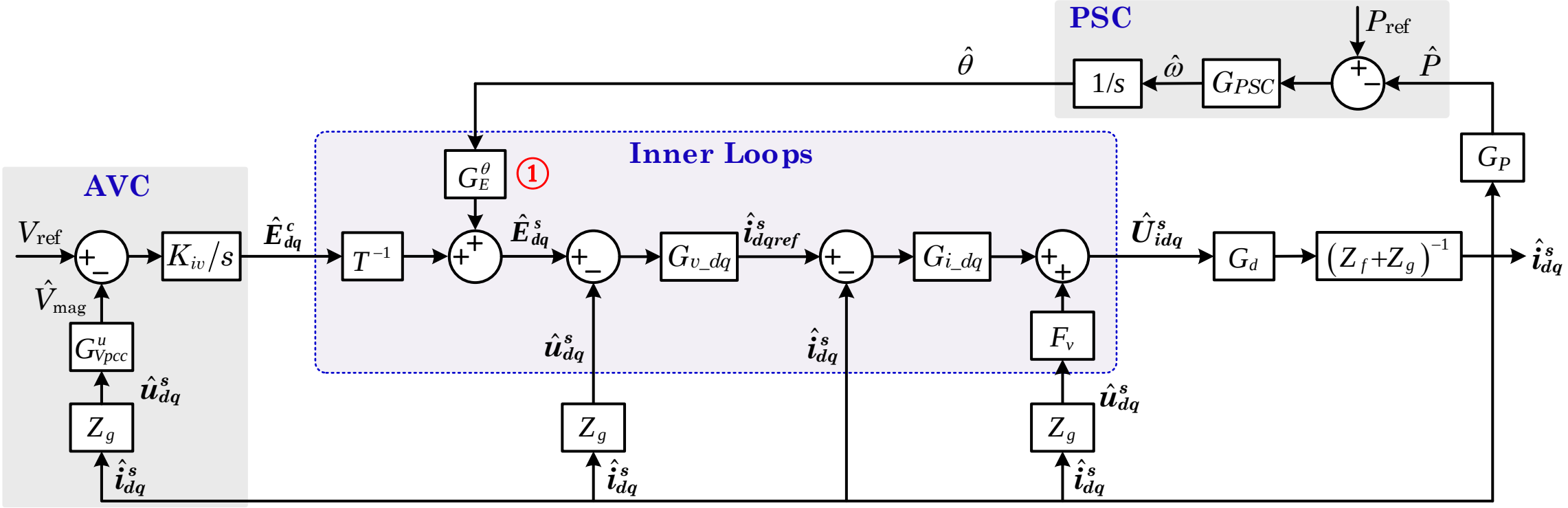

Fig. 17. Small-signal model of the GFM-VSC with $\alpha\beta$-frame realization of inner loops.

Fig. 14 illustrates complex torque profiles with different virtual inductance $L_v$, where $R_v$=0.1p.u., $k_{pi}$=2p.u., and PSC adopts a 3Hz LPF. It is shown that as $L_v$ transitions from 0.1p.u. to 0.2p.u., the net damping torque at the synchronous frequency shifts from negative to positive, implying the advantageous damping effect of $L_v$ on the SSO mode. Consequently, the $L_v$ can be utilized to mitigate the adverse damping effect of $R_v$ on the SSO mode. Moreover, $L_v$ has little effect on the SO-mode damping torque, which changes a little under the variation of $L_v$.

Fig. 15 shows complex torque profiles with different parameters of current control and voltage decoupling control. $L_v$=0.1p.u., $R_v$=0.1p.u., and PSC adopts a 3Hz LPF. Increasing current control P gain $k_{pi}$ and employing the voltage decoupling control ($F_v$=1) can enhance the damping torque of the SO mode, while slightly reducing the damping torque for the SSO mode. Yet, compared with virtual admittance control, the effect of current control and voltage decoupling control are insignificant.

In light of the impedance-based dynamic characterization and the complex torque coefficient-based interaction analysis, the damping effects of the inner-loop virtual admittance and current control, along with their design guidelines, are formulated as follows:

1) Current control with high bandwidth is recommended to alleviate its adverse interaction with virtual admittance control, thereby enhancing damping on the SO mode. Moreover, the voltage decoupling control should be employed to mitigate the adverse impact of interactions between virtual admittance control and current control.
2) Virtual resistance $R_v$ enhances the SO-mode damping but compromises the SSO-mode damping. Consequently, the design of $R_v$ should prioritize the damping effects on the SO mode, namely, by increasing the value of $R_v$ to mitigate the SO issues.
3) Virtual inductance $L_v$ enhances the SSO-mode damping and little impacts the SO-mode damping. Hence, the design of $L_v$ is dedicated to attenuating the SSO issue, through which the adverse effect of $R_v$ on SSO-mode damping can also be mitigated.

When $R_v$=0.1p.u., $L_v$≥0.2p.u., and $k_{pi}$=2p.u., the damping for both SO and SSO modes can be guaranteed, as shown in the blue line in Fig. 14.

Moreover, it is noted that parameter tuning of inner loops needs to be coordinated with the outer loop. Inner loops shape the output impedance and the electrical damping torque, providing a stable range for outer-loop parameters to be adjusted.

## V. Comparison of Different Reference-Frame Realizations of Inner Loops

The different reference-frame realizations of inner loops are also a critical concern to affect the inner-outer loop interactions and the system damping. This Section compares the control interactions with $\alpha\beta$-frame and $dq$-frame realizations of inner loops. The different coupling dynamics of those inner-loop realizations are illustrated via small-signal model; and their damping effects are characterized through the closed-loop poles.

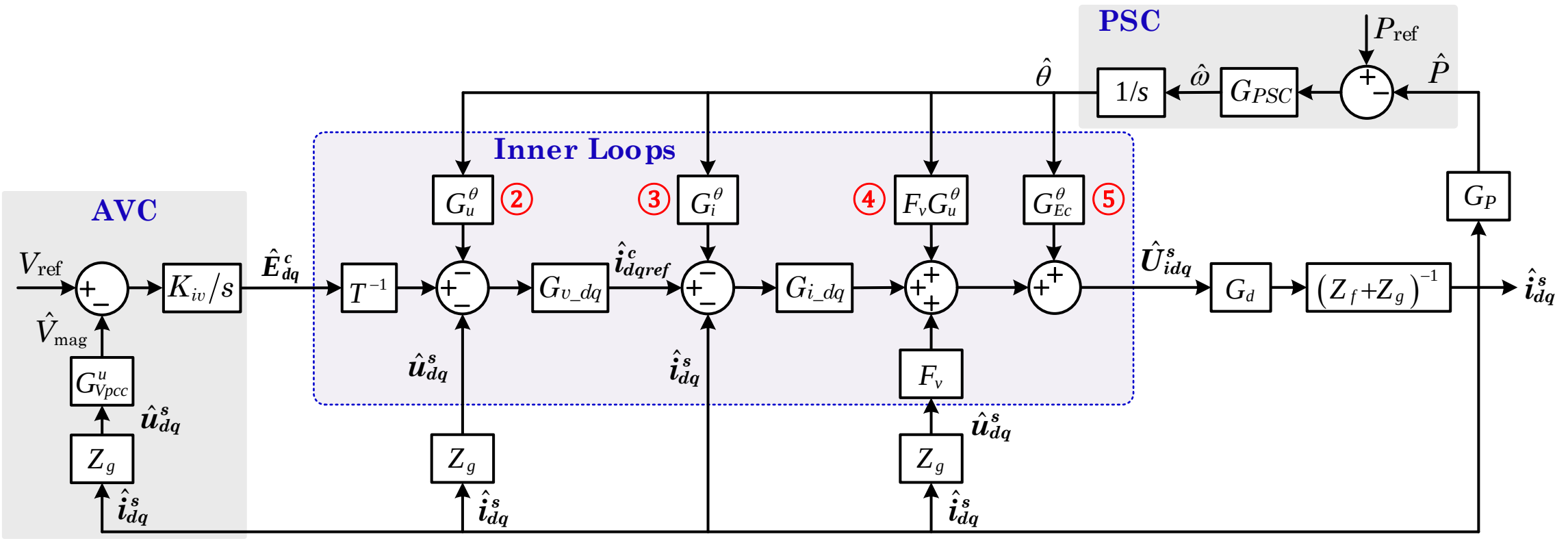

Fig. 18. Small-signal model of the GFM-VSC with *dq*-frame realization of inner loops.

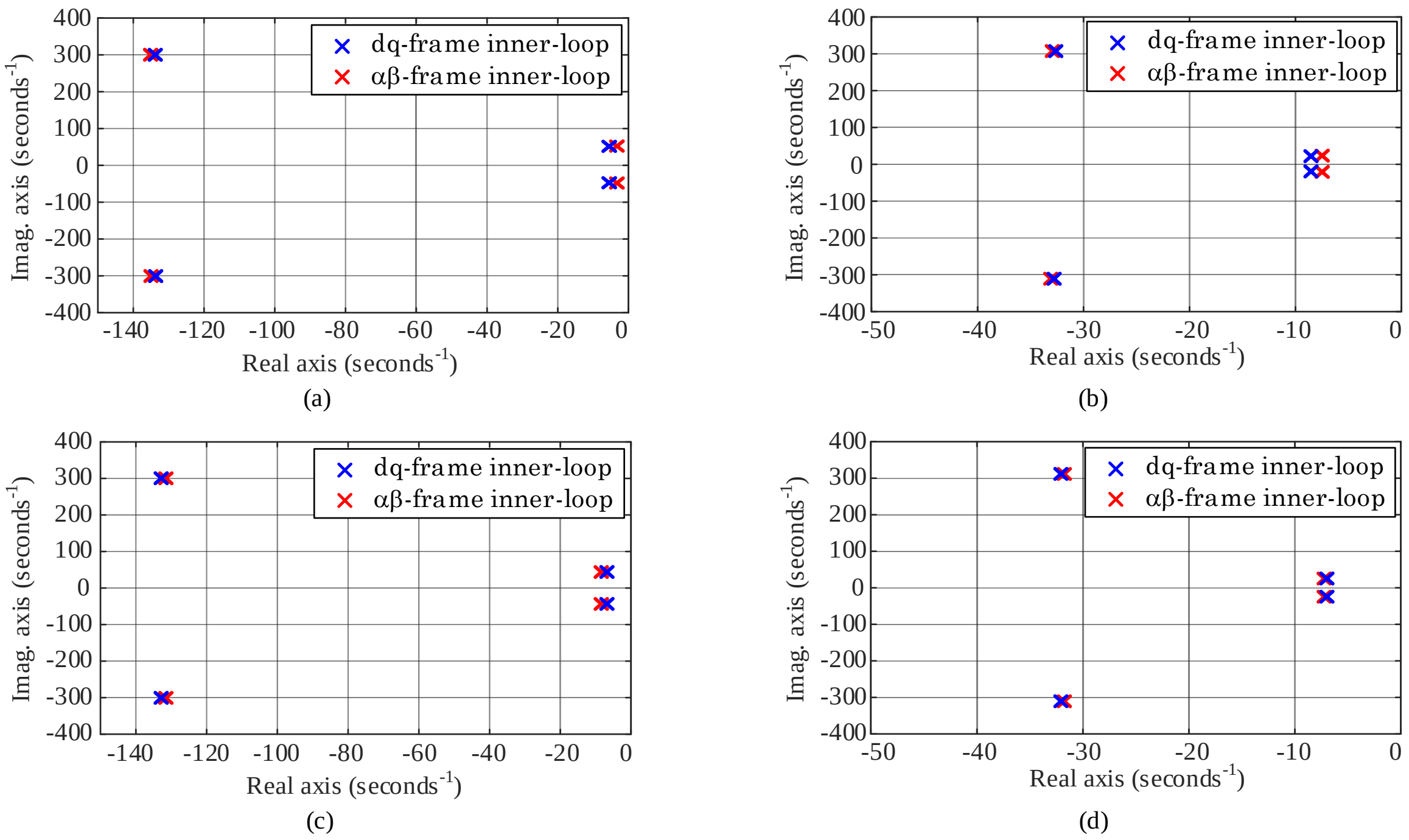

Fig. 19. Closed-loop poles of transfer function model. (a) Inverter mode with $L_g$=0.05 p.u. (b) Inverter mode with $L_g$=0.8 p.u. (c) Rectifier mode with $L_g$=0.05 p.u. (d) Rectifier mode with $L_g$=0.8 p.u.

### *A. Small-Single Model with αβ-Frame and dq-Frame Realizations of Inner Loops*

Fig. 16(a) shows the *αβ*-frame realization of inner loops. $\boldsymbol{E}_{dq} = \begin{bmatrix} E_{\text{ref}} & 0 \end{bmatrix}^T$ that is generated by outer loops is processed by *dq*-*αβ* transformation to generate the *αβ*-frame reference $\boldsymbol{E}_{\alpha\beta}$. The inner-loop control is implemented with *αβ*-frame variables $\boldsymbol{E}_{\alpha\beta}$, $\boldsymbol{V}_{\alpha\beta}$, and $\boldsymbol{i}_{\alpha\beta}$. Fig. 16(b) shows the *dq*-frame realization of inner loops. Differently, the inner-loop control is implemented with *αβ*-frame variables $\boldsymbol{E}_{dq}$, $\boldsymbol{V}_{dq}$, and $\boldsymbol{i}_{dq}$. It can be observed from the comparison that the phase angle used in *abc*-*dq* and *dq*-*abc* transformations results in different coupling dynamics with the outer-loop synchronization control.

Fig. 17 shows the small-signal model of the GFM-VSC with *αβ*-frame realization of inner loops. The detailed derivations of the matrix are presented in the Appendix [17], [34], [45]. The coupling loop ① denotes that the phase angle dynamic participates in the *dq*-*αβ* transformation of the EMF vector $\boldsymbol{E}_{\alpha\beta}$.

Fig. 18 shows the small-signal model of the GFM-VSC with *dq*-frame realization of inner loops. The detailed derivations of the matrix are presented in the Appendix [17], [34], [45]. The phase angle dynamic is coupled with the inner loop through four loops.

1) Loop ② denotes the *abc*-*dq* transformation of the PCC-voltage vector $\boldsymbol{V}_{pcc}$.
2) Loop ③ denotes the *abc*-*dq* transformation of the grid-current vector $\boldsymbol{i}_g$.
3) Loop ④ denotes the *abc*-*dq* transformation of the PCC-

voltage vector $\boldsymbol{V_{pcc}}$.

4) Loop ⑤ denotes the *dq-abc* transformation of the modulation-voltage vector $\boldsymbol{U_i}$.

## B. Comparative Analysis of Coupling Dynamics

The difference between *αβ*-frame and *dq*-frame realizations of inner loops lies in the coupling dynamics as depicted in loops ①–⑤. The matrix of loop ① in Fig. 17 and the matrix of loop ② in Fig. 18 are rewritten as

$$G_E^\theta = \begin{bmatrix} -E_{q0}^s \\ E_{d0}^s \end{bmatrix} \Leftrightarrow -G_u^\theta = \begin{bmatrix} -V_{q0}^s \\ V_{d0}^s \end{bmatrix}, \tag{19}$$

where $E_{d0}^s$ and $E_{q0}^s$ denote the steady-state operating point of the EMF vector. $V_{d0}^s$ and $V_{q0}^s$ represent the steady-state operating point of the PCC-voltage vector. It can be observed from the electrical circuit that the EMF vector and PCC-voltage vector are comparable, near to 1p.u. Therefore, loops ① and ② demonstrate comparable dynamics.

The loops ④ and ⑤ in Fig. 18 can be combined when the voltage decoupling control is enabled ($F_v$=1), given by

$$G_u^\theta + G_{Ui}^\theta = \begin{bmatrix} V_{q0}^s - U_{iq0}^s \\ U_{id0}^s - V_{d0}^s \end{bmatrix}, \tag{20}$$

where $U_{id0}^s$ and $U_{iq0}^s$ denote the steady-state operating point of the modulation-voltage vector. Since the modulation-voltage vector and the PCC-voltage vector are comparable in the electrical circuit, near 1p.u., the dynamics of loops ④ and ⑤ can be cancelled. Therefore, the difference between *αβ*-frame and *dq*-frame realizations of inner loops is primarily attributed to the coupling dynamics represented by loop ③, introduced by the *abc-dq* transformation of the grid-current vector $\boldsymbol{i_g}$. The matrix of loop ③, $G_i^\theta$, is represented by (41), which compromises the operating point of *dq*-frame current. The d-axis current $I_{d0}^s$ is a positive value under inverter mode and a negative value under rectifier mode. Therefore, the dynamic effect of the loop ③ is different under inverter and rectifier modes.

Fig. 19 shows the closed-loop poles with *αβ*- and *dq*-frame realizations of the inner loops. The results show that in the inverter operation mode, *dq*-frame realization demonstrates better damping for the SSO mode, while poorer damping for the SO mode. Conversely, in the rectifier operation mode, *dq*-frame realization exhibits poorer damping for the SSO mode, while better damping for the SO mode. However, this disparity is less apparent compared to the impact of control parameter variations. The inner loops can achieve favorable damping effects in both *αβ*- and *dq*-frame realizations by employing the robust parameter design presented in Section IV.

# VI. Simulation and Experimental Verification

Fig. 20 shows the experimental setup. The voltage source of ac grid is generated by the grid simulator. The VSC is connected to the grid through inductance filters and grid impedances. The control strategy is performed in the dSPACE-1007 platform.

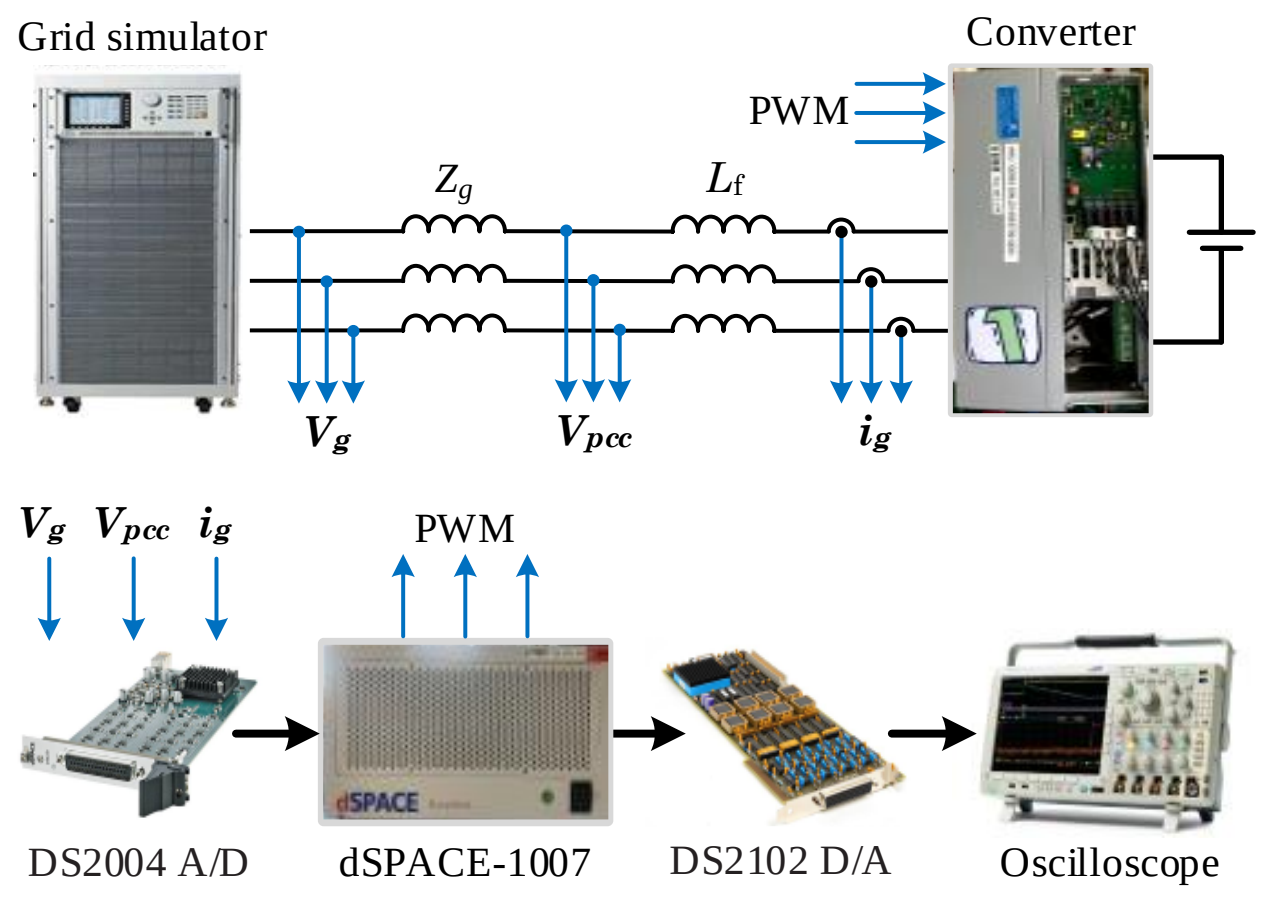

Fig. 20. Experimental setup.

TABLE II. Parameters for Simulation and Experimental Test

| Symbol | Description | Value (p.u.) |
|---|---|---|
| $P_n$ | Rated power | 3kW (1 p.u.) |
| $V_g$ | Rated voltage (L-G, RMS) | 110V/50Hz (1 p.u.) |
| $\omega_1$ | Nominal angular frequency | 100π (1 p.u.) |
| $L_g$ | Grid impedance | 2mH (0.05 p.u.) |
| $L_f$ | Filter inductance | 4mH (0.1 p.u.) |
| $V_1$ | Rated EMF magnitude | 110V (1 p.u.) |
| $K_{PSC}$ | PSC proportional gain | 0.1 $\omega_1/P_n$ (0.1 p.u.) |
| $K_{iv}$ | AVC integral gain | 50 (50 p.u.) |
| $L_v$ | Virtual inductance | 0.1-0.3 p.u. |
| $R_v$ | Virtual resistance | 0.05-0.12 p.u. |
| $k_{pi}$ | Current control proportional gain | 0.5-3 p.u. |
| $k_{ii}$ | Current control integral gain | 0.2 p.u. |
| $k_{ri}$ | Current control resonant gain | 0.4 p.u. |
| $T_d$ | Control time delay | 150 µs |

The main system constants and controller parameters are presented in Table II.

Fig. 21 shows the measured impedance profiles of inner loops, which are conducted via the frequency scan approach [46]. The impedance profiles are measured by disabling outer loops, thereby focusing only on the inner-loop dynamics. Fig. 21(a) compares the impedance profiles with and without the voltage decoupling control. The results demonstrate that the voltage decoupling control leads to an increase in the low-frequency resistance component. Fig. 21(b) presents impedance profiles with different $R_v$. It is shown that a large $R_v$ results in an increased resistance component in the low-frequency range while having little impact on the inductance component. Fig. 21(c) compares impedance profiles with different $L_v$. It is observed that a large $L_v$ corresponds to an increased inductance component with little impact on the resistance component. The measured impedance profiles validate the effectiveness of the proposed impedance models presented in Section III.

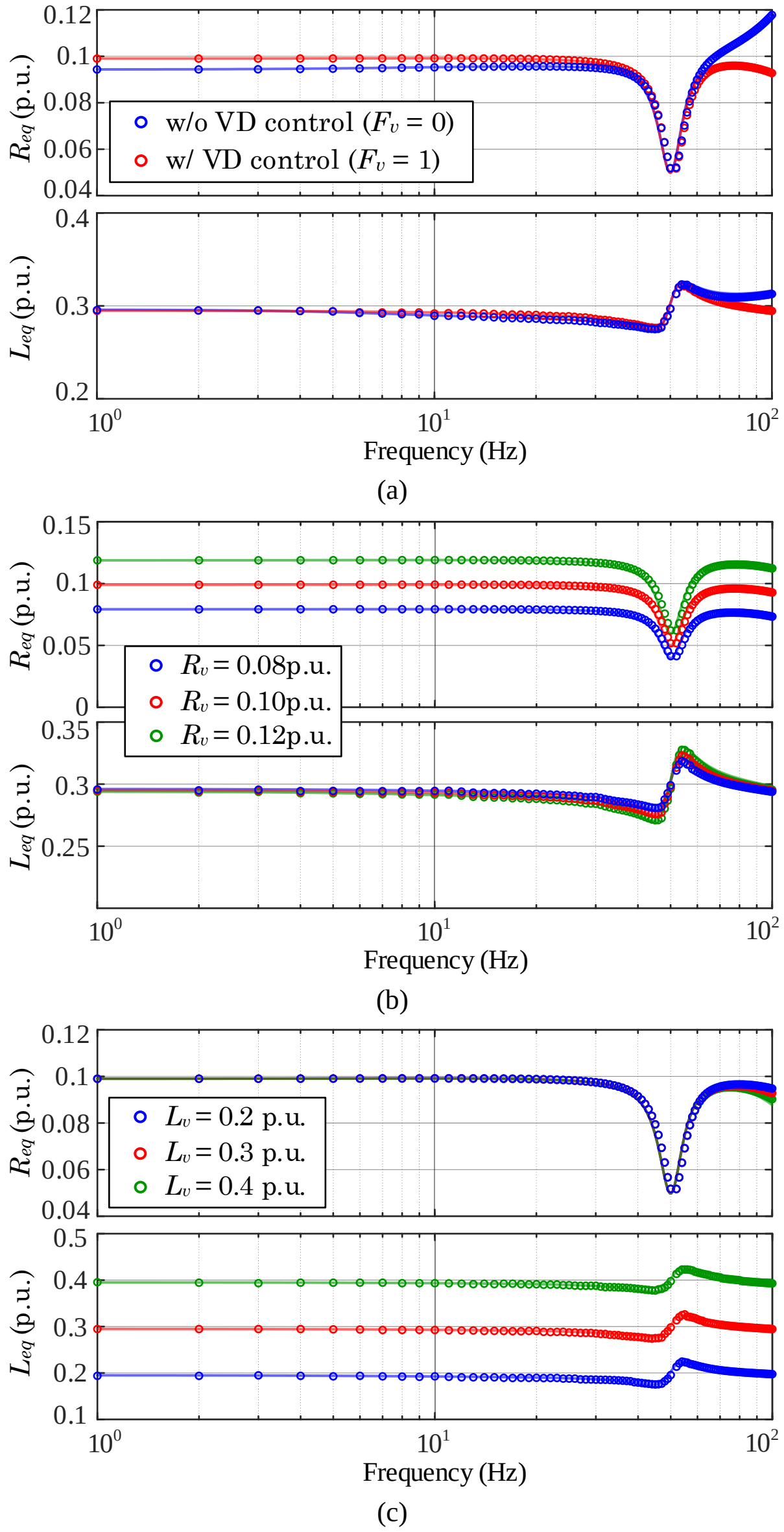

Fig. 21. Measured impedance profiles of inner-loop virtual admittance and current control. (a) Comparison with and without the voltage decoupling control. (b) With different $R_v$. (c) With different $L_v$.

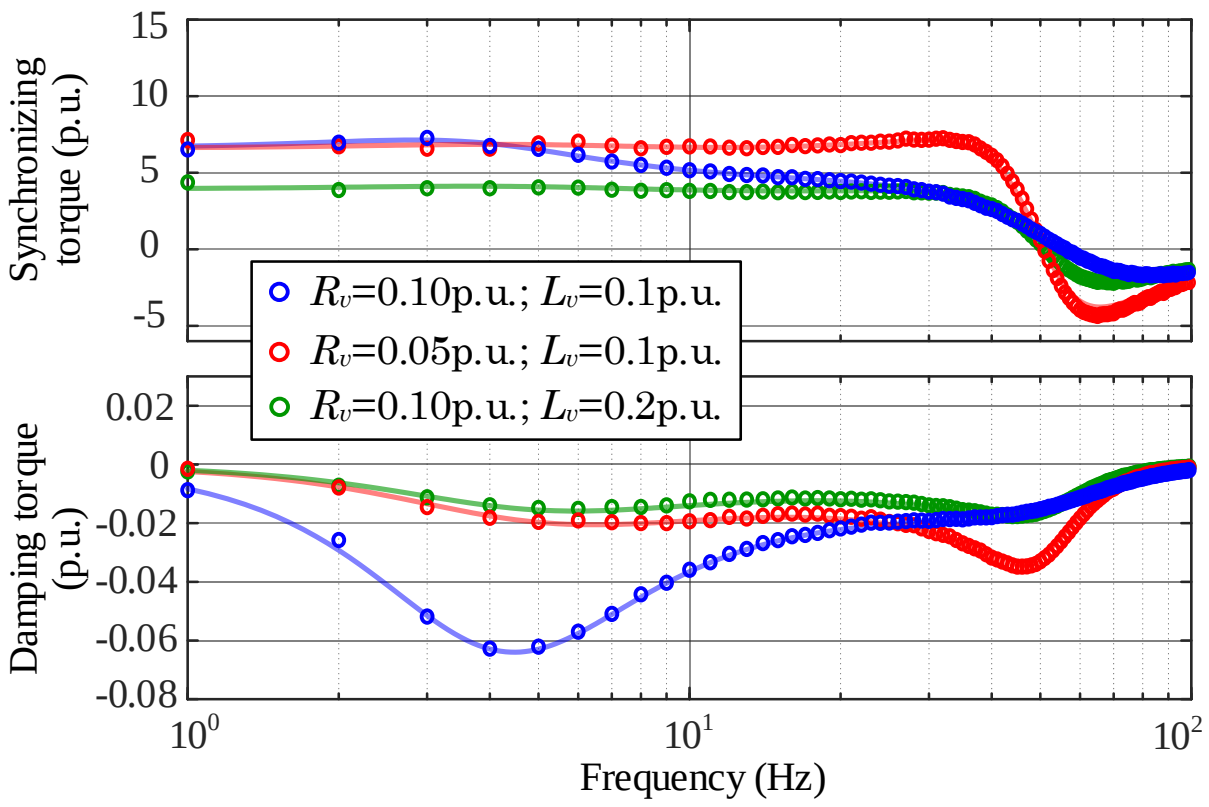

Fig. 22. Measured synchronizing and damping torques of inner-loop virtual admittance and current control.

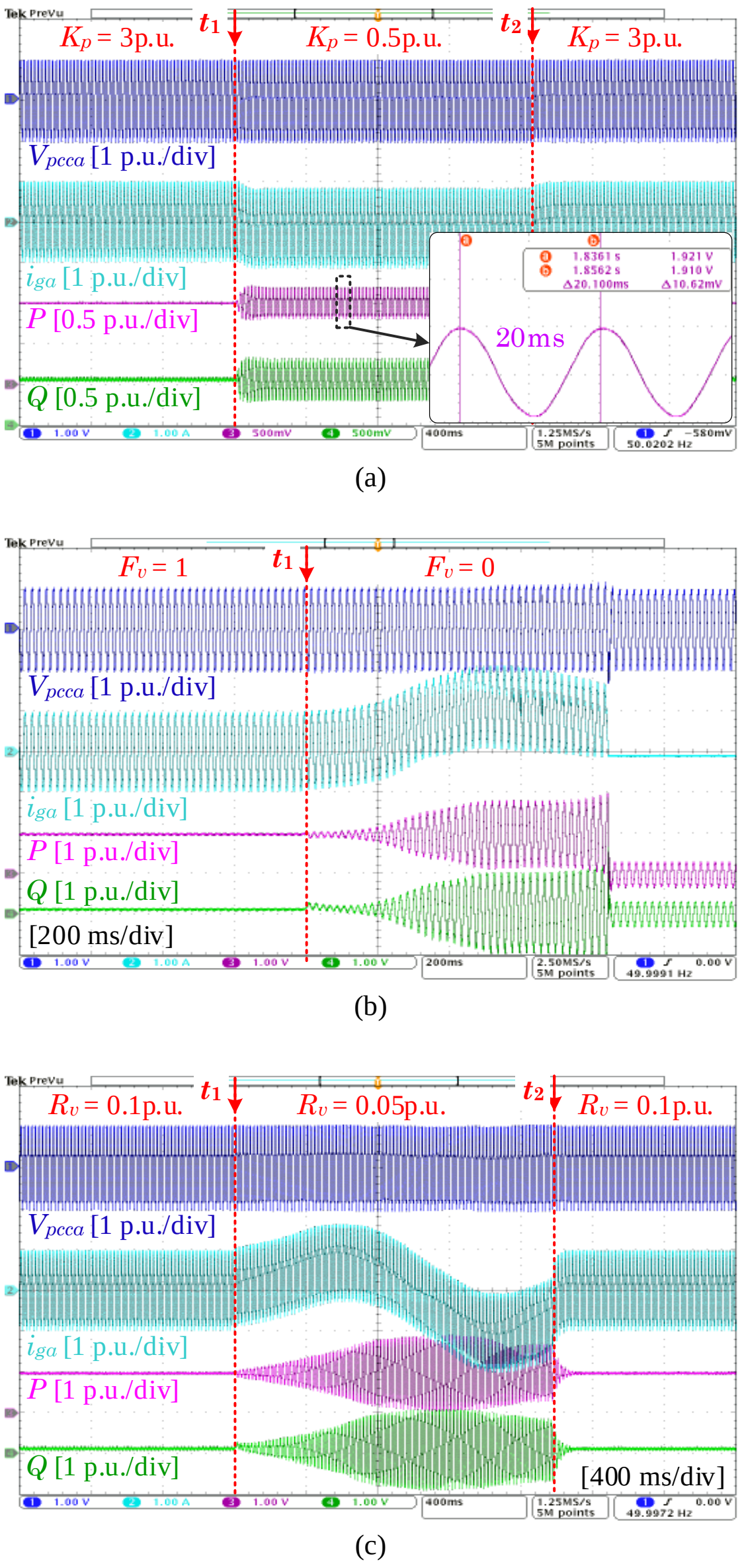

Fig. 23. Experimental waveform of SO-mode dynamics. (a) With different P gains of current control. (b) Comparison with and without the voltage decoupling control. (c) With different $R_v$.

Fig. 22 shows the measured complex torque coefficient of inner-loop virtual admittance and current control, corresponding to the transfer function from phase angle to active power $\boldsymbol{G_{P\theta}}$. The frequency scan approach is utilized for the measurement. The results indicate that increasing $R_v$ enhances damping torque for the SO mode but diminishes it for the SSO model. Conversely, a larger $L_v$ enhances the damping torque for the SSO mode but exhibits little influence on the damping torque of the SO mode. Those measured complex torque profiles confirm the validity of the theoretical complex torque models outlined in Section IV.

Fig. 23 shows experimental waveforms of SO-mode dynamics with virtual admittance and current control-based inner loops. Waveforms with different current control P gain $k_{pi}$ are compared in Fig. 23(a). It is shown that the system becomes unstable as $k_{pi}$ switches from 3p.u. to 0.5p.u. at $t_1$. A dc component is exhibited in the ac current waveform, and SO issues arise in active and reactive power waveforms. However, such issues are mitigated as $k_{pi}$ is switched back from 0.5p.u. to 3p.u. at $t_2$. It can be concluded that a large current control P gain has a beneficial damping effect on the SO mode, aligning with the theoretical analysis in Fig. 5.

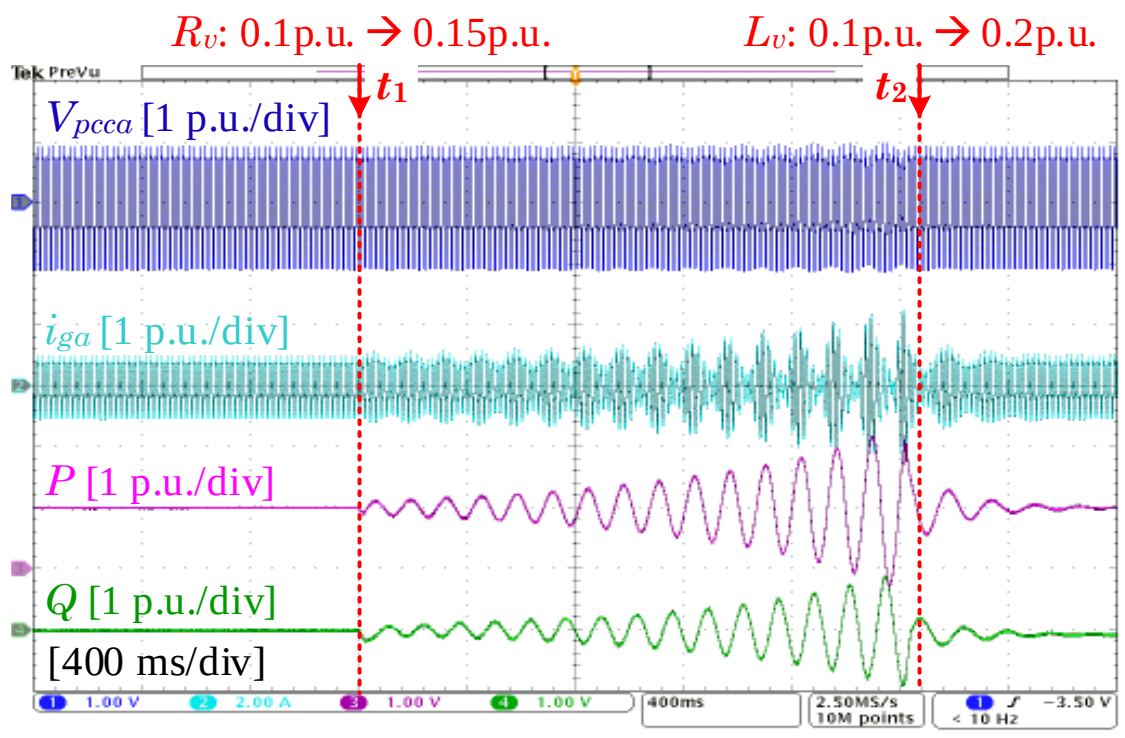

Fig. 24. Experimental waveform of SSO-mode dynamics.

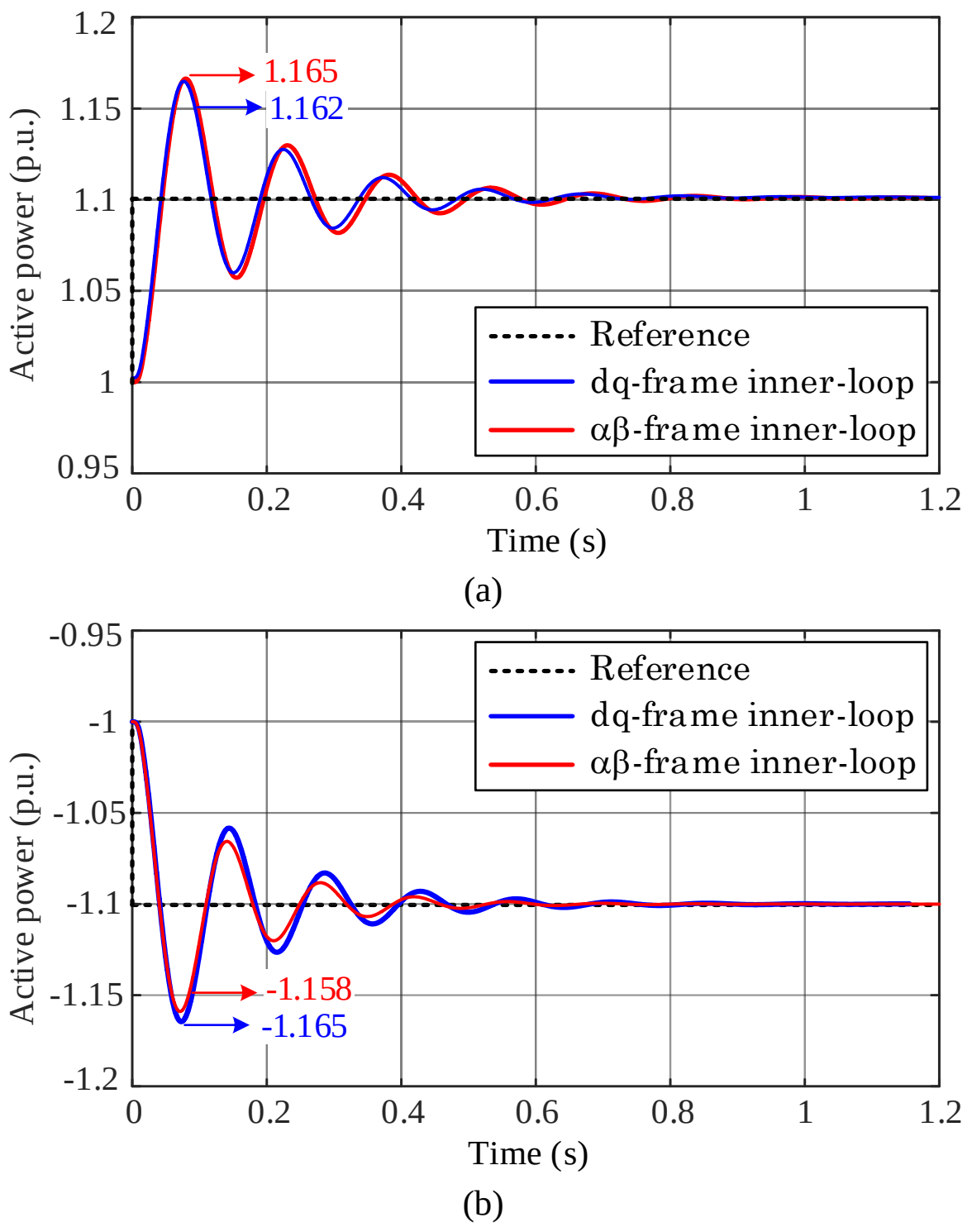

Fig. 25. Simulation step response of active power with $\alpha\beta$-frame and $dq$-frame realizations of inner loops. (a) Inverter mode. (b) Rectifier mode.

Fig. 23(b) shows the SO-mode dynamics with different virtual resistances $R_v$. Switching $R_v$ from 0.1p.u. to 0.05p.u. at $t_1$ results in the instability of the GFM-VSC, which is characterized by a dc component in the ac current waveform and SO issues in active and reactive power waveforms. Such SO issues are mitigated as $R_v$ is switched back from 0.05p.u. to 0.1p.u. at $t_2$. The results validate the beneficial damping effect of $R_v$ on the SO mode, which is consistent with the theoretical analysis in Fig. 7.

Fig. 23(c) compares the SO-mode dynamics with and without the voltage decoupling control. In this validation, the PSC proportional gain $K_{PSC}$ is selected as 0.17p.u., leading to a worse damping effect on the SO mode [34]. It can be observed that the GFM-VSC becomes unstable when disabling the voltage decoupling control, where a dc component is exhibited in the ac current waveform, and synchronous-frequency oscillations arise in active and reactive power waveforms. The results validate the beneficial damping effect of the voltage decoupling control on the SO mode., which is aligned with the theoretical analysis in Fig. 6.

Fig. 24 shows experimental waveforms of SSO-mode dynamics, where the GFM-VSC operates under the inverter mode with $L_g$=0.1p.u. It is shown that the system becomes unstable as $R_v$ changes from 0.1p.u. to 0.15p.u. at $t_1$. This instability manifests as low-frequency oscillations in voltage, current, and both active and reactive power waveforms. Subsequently, $L_v$ switches from 0.1p.u. to 0.2p.u. at $t_2$, mitigating those SSO issues. The results validate the adverse damping effect of $R_v$ and the beneficial damping effect of $L_v$ on the SSO-mode dynamics, aligning with the theoretical analysis in Fig. 13 and Fig. 14.

Fig. 25 shows the simulated step response of the active power. The $\alpha\beta$-frame and $dq$-frame realizations of inner loops are compared under both rectifier and inverter modes. When the active power reference is subject to a step change from 1p.u. to 1.1p.u., the GFM-VSC with $dq$-frame inner loops demonstrates a slightly smaller overshoot compared to $\alpha\beta$-frame inner loops. In contrast, during the rectifier mode, $\alpha\beta$-frame inner loops result in a slightly smaller active power overshoot than $dq$-frame inner loops. The comparative results validate the theoretical damping analysis for the SSO mode presented in Fig. 19.

## VII. Conclusion

This paper has developed an analytical method to assess the dynamics behaviour and damping effects of inner loops for GFM-VSCs.

1) An impedance model at the ac output of the GFM-VSC can uniformly characterize the behavior of various inner loops. The configuration and parameters of inner loops can shape the output impedance, subsequently influencing system dynamics.
2) The complex torque coefficient-based interaction analysis reveals that the net damping, contributed by outer loops, inner loops, and grid impedance, determines the system stability. Inner loops shape the electrical damping torque and subsequently influence the dynamics of both SO and SSO modes.
3) A study of virtual admittance and current control reveals that increasing the current control bandwidth and employing the voltage decoupling control are beneficial to enhancing low-frequency resistance. $R_v$ enhances the SO-mode damping but reduces the SSO-mode damping. In contrast, $L_v$ enhances the SSO-mode damping with little impact on the SO mode. Additionally, in the inverter mode, the *dq*-frame realization of inner loops shows slightly better damping for SSO mode and slightly reduced damping for SO mode compared to the *αβ*-frame realization, but this trend reverses in the rectifier mode.

## APPENDIX

### A. Modeling of Outer-Loop Dynamics

The synchronization dynamics result in two *dq*-frames, namely system and controller *dq*-frames. The PCC voltage aligns with the system *dq*-frame, with variables denoted by superscript "*s*". The EMF vector aligns with the controller *dq*-frame, with variables denoted by a superscript "*c*". Steady-state operating points are denoted by a subscript "$_0$". Small-signal perturbations are represented using a superscript "$\hat{}$".

The ac circuit dynamic is represented by

$$\left(\boldsymbol{V}_{\boldsymbol{dq0}}^{\boldsymbol{s}}+\hat{\boldsymbol{u}}_{\boldsymbol{dq}}^{\boldsymbol{s}}\right)-\boldsymbol{V}_{\boldsymbol{gdq0}}^{\boldsymbol{s}}=Z_g\left(\boldsymbol{I}_{\boldsymbol{dq0}}^{\boldsymbol{s}}+\hat{\boldsymbol{i}}_{\boldsymbol{dq}}^{\boldsymbol{s}}\right) \tag{21}$$

$$\hat{\boldsymbol{u}}_{\boldsymbol{dq}}^{\boldsymbol{s}}=Z_g\cdot\hat{\boldsymbol{i}}_{\boldsymbol{dq}}^{\boldsymbol{s}} \tag{22}$$

The grid impedance is given by

$$Z_g=\begin{bmatrix} sL_g & -\omega_1 L_g \\ \omega_1 L_g & sL_g \end{bmatrix} \tag{23}$$

The active power dynamics are represented by

$$\hat{P}=\underbrace{\begin{bmatrix} I_{d0}^s & I_{q0}^s \end{bmatrix}}_{G_P^u}\cdot\hat{\boldsymbol{u}}_{\boldsymbol{dq}}^{\boldsymbol{s}}+\underbrace{\begin{bmatrix} V_{d0}^s & V_{q0}^s \end{bmatrix}}_{G_P^i}\cdot\hat{\boldsymbol{i}}_{\boldsymbol{dq}}^{\boldsymbol{s}} \tag{24}$$

$$\hat{P}=G_P\cdot\hat{\boldsymbol{i}}_{\boldsymbol{dq}}^{\boldsymbol{s}},\quad G_P=G_P^i+G_P^u\cdot Z_g \tag{25}$$

The synchronization dynamics are given by

$$\hat{\theta}=\frac{G_{PSC}}{s}\cdot\left(P_{\text{ref}}-\hat{P}\right). \tag{26}$$

The PCC voltage magnitude dynamics are given by

$$\hat{V}_{\text{mag}}=G_{Vpcc}^u\cdot\hat{\boldsymbol{u}}_{\boldsymbol{dq}}^{\boldsymbol{s}},\quad G_{Vpcc}^u=\frac{\begin{bmatrix} V_{d0}^s & V_{q0}^s \end{bmatrix}}{\sqrt{\left(V_{d0}^s\right)^2+\left(V_{q0}^s\right)^2}} \tag{27}$$

The dynamic of AVC is given by

$$\hat{E}=\frac{K_{iv}}{s}\cdot\left(V_{\text{ref}}-\hat{V}_{\text{mag}}\right). \tag{28}$$

### B. Modeling of αβ-Frame Realization of Inner Loops

For the *αβ*-frame realization of inner loops, the EMF vector is first generated through the *dq-αβ* transformation, given by

$$\boldsymbol{E}_{\boldsymbol{dq0}}^{\boldsymbol{s}}+\hat{\boldsymbol{E}}_{\boldsymbol{dq}}^{\boldsymbol{s}}=e^{j\left(\theta_0+\hat{\theta}\right)}\cdot\left(\boldsymbol{E}_{\boldsymbol{dq0}}^{\boldsymbol{c}}+\hat{\boldsymbol{E}}_{\boldsymbol{dq}}^{\boldsymbol{c}}\right), \tag{29}$$

$$\hat{\boldsymbol{E}}_{\boldsymbol{dq}}^{\boldsymbol{s}}=T^{-1}\cdot\hat{\boldsymbol{E}}_{\boldsymbol{dq}}^{\boldsymbol{c}}+G_E^\theta\cdot\hat{\theta},\quad G_E^\theta=\begin{bmatrix} -E_{q0}^s \\ E_{d0}^s \end{bmatrix}, \tag{30}$$

$$T^{-1}=e^{j\theta_0}=\begin{bmatrix} \cos\theta_0 & -\sin\theta_0 \\ \sin\theta_0 & \cos\theta_0 \end{bmatrix} \tag{31}$$

The control laws of inner loops are given by

$$\hat{\boldsymbol{i}}_{\boldsymbol{dqref}}^{\boldsymbol{s}}=G_{v_dq}\cdot\left(\hat{\boldsymbol{E}}_{\boldsymbol{dq}}^{\boldsymbol{s}}-\hat{\boldsymbol{u}}_{\boldsymbol{dq}}^{\boldsymbol{s}}\right), \tag{32}$$

$$\hat{\boldsymbol{U}}_{\boldsymbol{idq}}^{\boldsymbol{s}}=G_{i_dq}\cdot\left(\hat{\boldsymbol{i}}_{\boldsymbol{dqref}}^{\boldsymbol{s}}-\hat{\boldsymbol{i}}_{\boldsymbol{dq}}^{\boldsymbol{s}}\right)+F_v\cdot\hat{\boldsymbol{u}}_{\boldsymbol{dq}}^{\boldsymbol{s}} \tag{33}$$

$$\begin{aligned} G_{v_dq}(s)&=G_{v_\alpha\beta}(s+j\omega_1) \\ G_{i_dq}(s)&=G_{i_\alpha\beta}(s+j\omega_1) \end{aligned} \tag{34}$$

The open-loop and closed-loop transfer functions, from the reference to the output of active power, are given by

$$\begin{cases} T_{open_\alpha\beta}=G_P\left(Z_{\text{f}}+Z_g-\boldsymbol{T_1}\right)^{-1}G_dG_{i_dq}G_{v_dq}G_E^\theta\dfrac{G_{PSC}}{s} \\ \boldsymbol{T_1}=\begin{pmatrix} G_dF_vZ_g-G_dG_{i_dq}-G_dG_{i_dq}G_{v_dq}Z_g \\ -G_dG_{i_dq}G_{v_dq}T^{-1}K_{iv}G_{Vpcc}^uZ_g/s \end{pmatrix}^{-1} \end{cases} \tag{35}$$

$$T_{closed_\alpha\beta}=\frac{\hat{P}_{\text{pcc}}}{\hat{P}_{\text{ref}}}(s)=\frac{T_{open_\alpha\beta}}{1+T_{open_\alpha\beta}} \tag{36}$$

### C. Modeling of dq-Frame Realization of Inner Loops

For the *dq*-frame realization of inner loops, the *abc-dq* transformation of PCC voltage is represented by

$$\boldsymbol{U}_{\boldsymbol{dq0}}^{\boldsymbol{c}}+\hat{\boldsymbol{u}}_{\boldsymbol{dq}}^{\boldsymbol{c}}=e^{-j\left(\theta_0+\hat{\theta}\right)}\cdot\left(\boldsymbol{U}_{\boldsymbol{dq0}}^{\boldsymbol{s}}+\hat{\boldsymbol{u}}_{\boldsymbol{dq}}^{\boldsymbol{s}}\right) \tag{37}$$

$$\hat{\boldsymbol{u}}_{\boldsymbol{dq}}^{\boldsymbol{c}}=T\cdot\left(\hat{\boldsymbol{u}}_{\boldsymbol{dq}}^{\boldsymbol{s}}+G_u^\theta\cdot\hat{\theta}\right),\quad G_u^\theta=\begin{bmatrix} V_{q0}^s \\ -V_{d0}^s \end{bmatrix} \tag{38}$$

$$T=e^{-j\theta_0}=\begin{bmatrix} \cos\theta_0 & \sin\theta_0 \\ -\sin\theta_0 & \cos\theta_0 \end{bmatrix} \tag{39}$$

The *abc-dq* transformation of grid current is represented by

$$\boldsymbol{I}_{\boldsymbol{dq0}}^{\boldsymbol{c}}+\hat{\boldsymbol{i}}_{\boldsymbol{dq}}^{\boldsymbol{c}}=e^{-j\left(\theta_0+\hat{\theta}\right)}\cdot\left(\boldsymbol{I}_{\boldsymbol{dq0}}^{\boldsymbol{s}}+\hat{\boldsymbol{i}}_{\boldsymbol{dq}}^{\boldsymbol{s}}\right) \tag{40}$$

$$\hat{\boldsymbol{i}}_{\boldsymbol{dq}}^{\boldsymbol{c}}=T\cdot\left(\hat{\boldsymbol{i}}_{\boldsymbol{dq}}^{\boldsymbol{s}}+G_i^\theta\cdot\hat{\theta}\right),\quad G_i^\theta=\begin{bmatrix} I_{q0}^s \\ -I_{d0}^s \end{bmatrix} \tag{41}$$

The small-signal representations of control laws are given by

$$\hat{\boldsymbol{i}}_{\boldsymbol{dqref}}^{\boldsymbol{c}}=G_{v_dq}\cdot\left(\hat{\boldsymbol{E}}_{\boldsymbol{dq}}^{\boldsymbol{c}}-\hat{\boldsymbol{u}}_{\boldsymbol{dq}}^{\boldsymbol{c}}\right), \tag{42}$$

$$\hat{\boldsymbol{U}}_{\boldsymbol{idq}}^{\boldsymbol{c}}=G_{i_dq}\cdot\left(\hat{\boldsymbol{i}}_{\boldsymbol{dqref}}^{\boldsymbol{c}}-\hat{\boldsymbol{i}}_{\boldsymbol{dq}}^{\boldsymbol{c}}\right)+F_v\cdot\hat{\boldsymbol{u}}_{\boldsymbol{dq}}^{\boldsymbol{c}} \tag{43}$$

The *dq-abc* transformation of modulation voltage is represented by

$$\boldsymbol{U}_{\boldsymbol{idq0}}^{\boldsymbol{s}}+\hat{\boldsymbol{U}}_{\boldsymbol{idq}}^{\boldsymbol{s}}=e^{j\left(\theta_0+\hat{\theta}\right)}\cdot\left(\boldsymbol{U}_{\boldsymbol{idq0}}^{\boldsymbol{c}}+\hat{\boldsymbol{U}}_{\boldsymbol{idq}}^{\boldsymbol{c}}\right) \tag{44}$$

$$\hat{\boldsymbol{U}}_{\boldsymbol{idq}}^{\boldsymbol{s}} = T^{-1} \cdot \hat{\boldsymbol{U}}_{\boldsymbol{idq}}^{\boldsymbol{c}} + G_{Ui}^{\theta} \cdot \hat{\theta}, \quad G_{Ui}^{\theta} = \begin{bmatrix} -U_{iq0}^{s} \\ U_{id0}^{s} \end{bmatrix} \tag{45}$$

The open-loop and closed-loop transfer functions, from the reference to the output of active power, are given by

$$\begin{cases} T_{open_\alpha\beta} = G_P \left( Z_{\mathrm{f}} + Z_g - \boldsymbol{T_2} \right)^{-1} G_d \boldsymbol{T_3} G_{PSC} / s \\ \boldsymbol{T_2} = \begin{pmatrix} G_d F_v Z_g - G_d G_{i_dq} - G_d G_{i_dq} G_{v_dq} Z_g \\ -G_d G_{i_dq} G_{v_dq} T^{-1} K_{iv} G_{Vpcc}^{u} Z_g / s \end{pmatrix}^{-1} \\ \boldsymbol{T_3} = G_{Ui}^{\theta} + F_v G_u^{\theta} - G_{i_dq} G_i^{\theta} - G_{i_dq} G_{v_dq} G_u^{\theta} \end{cases} \tag{46}$$

$$T_{closed_dq} = \frac{\hat{P}_{\mathrm{pcc}}}{\hat{P}_{\mathrm{ref}}}(s) = \frac{T_{open_dq}}{1 + T_{open_dq}} \tag{47}$$